\documentclass[aps,prd,a4paper,onecolumn,amsmath,showpacs,superscriptaddress,nofootinbib,preprintnumbers,notitlepage]{revtex4-1}
 
\usepackage{verbatim}
\usepackage[T1]{fontenc}
\usepackage[utf8]{inputenc}
\usepackage[american]{babel}
\usepackage{graphicx,subcaption,caption}
\usepackage{epsfig}
\usepackage{graphicx,subcaption,caption}
\usepackage{pgfplots}
\usepackage{booktabs}
\usepackage{multirow}
\usepackage{dcolumn}
\usepackage{amsmath}
\usepackage{mathtools}
\usepackage{amsfonts}
\usepackage{amssymb}
\usepackage{epstopdf}
\usepackage{bm}
\usepackage{siunitx}
\usepackage{braket}
\usepackage{enumitem}
\usepackage{soul}
\usepackage{color}
\usepackage{amsmath}
\DeclareMathOperator{\sech}{sech}

\DeclareMathOperator{\arcsinh}{arcsinh}
\DeclareMathOperator{\arctanh}{arctanh}

\usepackage{transparent}
\usepackage{pifont}

\definecolor{navyblue}{rgb}{0.0, 0.0, 0.5}
\definecolor{royalblue}{rgb}{0.25, 0.41, 0.88}
\definecolor{cadmiumgreen}{rgb}{0.0, 0.42, 0.24}
\definecolor{blue-violet}{rgb}{0.54, 0.17, 0.89}
\definecolor{darkviolet}{rgb}{0.58, 0.0, 0.83}
\definecolor{orange(colorwheel)}{rgb}{1.0, 0.5, 0.0}

\usepackage{hyperref}
\hypersetup{
    colorlinks=true, % false: boxed links; true: colored links
    linkcolor=royalblue, % color of internal links (change box color with linkbordercolor)
    citecolor=magenta}

\newcommand\be{\begin{equation}}
\newcommand\ee{\end{equation}}

\newcommand\bea{\begin{eqnarray}}
\newcommand\eea{\end{eqnarray}}

\usepackage{booktabs}
\usepackage{multirow}
\usepackage{dcolumn}
\usepackage{colortbl}

\definecolor{magenta(process)}{rgb}{1.0, 0.0, 0.56}

\definecolor{darkspringgreen}{rgb}{0.09, 0.45, 0.27}

\definecolor{royalblue(web)}{rgb}{0.25, 0.41, 0.88}

\begin{document}

\title{Galilean constant-roll inflation}

\author{Ram\'{o}n Herrera}
\email{ramon.herrera@pucv.cl}
\affiliation{Instituto de F\'{\i}sica, Pontificia Universidad Cat\'{o}lica de Valpara\'{\i}so, Avenida Brasil 2950, Casilla 4059, Valpara\'{\i}so, Chile
}

\author{Mehdi Shokri}
\email{mehdishokriphysics@gmail.com}
\affiliation{Department of Physics, University of Tehran, North Karegar Ave., Tehran 14395-547, Iran}
\affiliation{School of Physics, Damghan University, P. O. Box 3671641167, Damghan, Iran}
\affiliation{Canadian Quantum Research Center 204-3002 32 Avenue Vernon, British Columbia V1T 2L7 Canada}

\author{Jafar Sadeghi}
\email{pouriya@ipm.ir}
\affiliation{Department of Physics, University of Mazandaran, P. O. Box 47416-95447, Babolsar, Iran}

\preprint{}
\begin{abstract}
 The constant-roll inflation in the context of Galilean inflation 
 or G-inflation is analyzed. By considering some coupling function $G(\varphi,\chi)$ associated with the model of G-inflation, we find different inflationary solutions in the context of the constant roll scenario.
 In order to present an  analytical discussion, 
 we 
 work with two specific cases of the general function $G(\varphi,\chi)\propto g(\varphi)\,\chi^n$, e.g. i) $G(\varphi,\chi)\propto\,\varphi$ when $g(\varphi)=\varphi$ and $n=0$ and ii) $G(\varphi,\chi)\propto\,\sqrt{\chi}$ when $g(\varphi)=$ constant and $n=1/2$. Also, we introduce a new function $G(\varphi,\chi)$ associated with both variables $\varphi$ and $\chi$. We
reconstruct the potential of the scalar field for the considered cases of the function $G$ in the context of the constant-roll approach and then we study the corresponding cosmological perturbations of the model. Eventually, we use the recent observations datasets in order to constrain the parameter-space of the model.
%\\\\
%\\{\bf PACS:} 98.80.Cq; 98.80.$-$K.
%\\{\bf Keywords}:  G-inflation; Constant-roll approach; Cosmic Microwave Background. 
\end{abstract}
\maketitle

\section{Introduction}

Today cosmic inflation has been known as an unavoidable part of modern cosmology capable to remove the Hot Big Bang defects \textit{i.e.}, flatness, horizon and monopole problems by considering a rapid and enormous expansion at the early time. Moreover, inflation is the main responsible for the structure formation of the universe on large scale through generating scalar perturbation. Tensor perturbations produced during the inflationary era also can be traced by monitoring the polarization anisotropies of Cosmic Microwave Background (CMB) photons created by the primordial gravitational waves \cite{Guth,Linde:1981my,Albrecht:1982wi,Lyth:1998xn}. From the slow-roll inflationary viewpoint, a single scalar field, the so-called \textit{inflaton}, is the dominant component of the universe rolling down slowly from the peak of the potential at the start of inflation to the bottom at the end of inflation. Then, inflaton decays to the particles at the last step of inflation through the reheating procedure \cite{Kofman2,Shtanov}. A wide range of inflationary papers has investigated the properties of the single field models by comparing them with the CMB anisotropies observations. Consequently, some of the single models failed, while some of them are still in good agreement with the satellite releases \cite{martin,staro,barrow,kallosh1}. Despite the advantages of the single field models, they suffer from the lack of non-Gaussinaity in the spectrum through the uncorrelated modes \cite{Chen}. This can be problematic when our future observations show the non-Gaussianity in the perturbations spectrum. Hence, the constant-roll idea is introduced by considering a constant rate of rolling for the scalar field during the inflationary era as
\begin{equation}
\ddot{\varphi}=-(3+\alpha)H\dot{\varphi},
\label{1}
\end{equation}
where $\alpha$ is a non-zero parameter \cite{martin2,Motohashi1,Motohashi2}. Going beyond the slow-roll approximation also can be addressed in the ultra slow-roll (USR) regime where we deal with a non-negligible $\ddot{\varphi}$ in the Klein-Gordon equation as $\ddot{\varphi}=-3H\dot{\varphi}$. The USR model shows a finite value for the non-decaying mode of curvature perturbations \cite{Inoue}. Also, the USR solutions are situated in the non-attractor phase of inflation but sometimes are characterized by an attractor-like behavior dynamically \cite{Pattison}. Although the USR model predicts a large $\eta$, it is not able to solve the $\eta$ problem introduced in supergravity for the hybrid inflationary models \cite{Kinney}. Besides the USR models, a fast-roll model can be introduced by considering a fast-rolling stage at the first steps of inflation \cite{Contaldi,Lello,Hazra}. Recently, constant-roll inflation has been considered remarkably among plenty of inflationary papers \cite{Odintsov,Nojiri,Motohashi9,Awad:2017ign,Cicciarella,Anguelova,Ito,Ghersi,Lin,Micu,Kamali,Oliveros,Motohashi3,lq,lqq,ll0,ll1,shokriii, Mohammadi:2022tmk}.

Regarding different inflationary models, here we deal with a special class of models, named in the literature as Galilean inflationary models or merely G-inflation. Furthermore, its generalization, called  generalized G-inflation (or G$^2$-inflation) which also corresponds to a subclass of the Horndeski theory \cite{Horndeski:1974wa}, was studied in Ref. \cite{Kobayashi:2011nu}. 

In this framework, the Galilean action assumes an additional term $G(\varphi,\chi)\Box\varphi$ to the standard action, where $G(\varphi,\chi)$ corresponds to an arbitrary function of the scalar field $\varphi$ and of the quantity $\chi$ defined as $\chi=-g^{\mu\nu}\partial_{\mu}\varphi\partial_{\nu}\varphi/2$. Note that this new term does not change the speed gravitational waves being equivalent to the speed of light \cite{Kobayashi:2010cm,Deffayet:2010qz}. Also, we mention that the study of these models deserves a careful analysis to prevent the appearance of instabilities and have successful inflation \cite{Kobayashi:2011nu,Kobayashi:2010cm}, as well as an appropriate scenario of reheating \cite{BazrafshanMoghaddam:2016tdk}. In this context, different inflationary models have been studied in the frame of Galilean inflation. In particular, the model of Higgs G-inflation which considers the slow roll regime and takes some effective potentials associated with the Higgs field was studied in \cite{Kamada:2010qe}. Also, the model of USR G-inflation was studied in Ref.\cite{Hirano:2016gmv}, in which the effective potential associated with the scalar field is considered a constant. Moreover, the situation in which the scalar potential is of the power-law type was analyzed in Ref.\cite{Unnikrishnan:2013rka}. Concerning the Primordial non-Gaussianity in the context of G-inflation was developed in \cite{Kobayashi:2011pc}, see also \cite{Zhang:2020uek}. In the framework of warm G-inflation the dynamics and its thermal fluctuations were developed in Ref.\cite{Herrera:2017qux}, see also \cite{Herrera:2018wan,Motaharfar:2018mni} for other warm G-inflationary models. 

The main goal of the present manuscript is to study the Galilean inflationary model in the context of the constant-roll idea. We investigate the parameter space of the G-inflation and the cosmological perturbations under the constant-roll concept. In this way, we analyze how the function $G(\varphi,\chi)$,  modifies the background variables, such as the scalar potential $V(\varphi)$, the Hubble parameter and  also the constraints on the parameters from  cosmological perturbations. To fulfill this goal, we assume a coupling for the function $G(\varphi,\chi)$ given by  $G(\varphi,\chi)\propto g(\phi)\chi^n$, see Ref. \cite{Herrera:2018wan}. In order to obtain analytical solutions for the constant roll inflation, we study the specific case in which $g(\varphi)=\varphi$ and $n=0$, as well the situation in which $g(\varphi)=$ constant and $n=1$. Besides the mentioned cases, we study the model for a particular form of $G$ as a mixture of two variables $\varphi$ and $\chi$ which is compatible with the concept of the constant-roll approach.

The above discussion motivates us to organize this paper as follows. In the next section, we present a brief review of the background equations together with the cosmological perturbations in the framework of the
G-inflation. In Sect. \ref{GCR}, we consider the constant-roll condition and we analyze three specific cases for the coupling function $G(\varphi,\chi)$. In particular, we assume the case in which the function $G(\varphi,\chi)\propto\varphi$ and also we study the specific case in which the coupling function $G(\varphi,\chi)\propto \sqrt{\chi}$ in the context of G-constant roll inflation. Besides we analyze a coupling function $G(\varphi,\chi)$ that depends of both variables, the scalar field as well of  its kinetic energy. Additionally, in this section we find the different constraints on the parameter-space,  using the cosmological perturbations and  constrained from Planck data. Also,  we summarize our findings and present our conclusions in Sect. \ref{Con}. In the following, we chose units so that $c=\hbar= \kappa^2= 8\pi G=1.$

\section{G-inflation}

In this section, we present a brief review of the background equations and the cosmological perturbations associated with the G-inflationary model. We start with the 4-dimensional action $S$ in the framework of the  Galilean model in which
  
\begin{equation}
S=\int{d^{4}x\sqrt{-g}\bigg(\frac{R}{2}+K(\varphi,\chi)-G(\varphi,\chi)\displaystyle \Box\varphi}\bigg),
\label{2}   
\end{equation}
where $g$ is the determinant of the metric $g_{\mu\nu}$, $R=g^{\mu\nu}R_{\mu\nu}$ is the Ricci scalar and as before, $\varphi$ is the scalar field and $\chi=-g^{\mu\nu}\partial_{\mu}\varphi\partial_{\nu}\varphi/2$. Also, the quantities $K$ and $G$ are arbitrary functions of $\chi$ and $\varphi$.  By varying the action (\ref{2}) with respect to the metric and then using a spatially flat Friedmann-Robertson-Walker (FRW) metric 
%describing an isotropic and homogeneous universe, 
and considering a homogeneous scalar field $\varphi=\varphi(t)$,
the modified dynamical equations can be written as
\begin{equation}
3H^{2}+K+\dot{\varphi}^{2}(G_{\varphi}-K_{\chi})-3HG_{\chi}\dot{\varphi}^{3}=0,
\label{3}
\end{equation}
and
\begin{equation}
2\dot{H}+3H^{2}+K-\dot{\varphi}^{2}(G_{\varphi}+G_{\chi}\ddot{\varphi})=0.
\label{4}
\end{equation}
Similarly, by varying the action (\ref{2}) with respect to the scalar field $\varphi$, the modified Klein-Gordon equation becomes
%is obtained as
\begin{eqnarray}
(\ddot{\varphi}+3H\dot{\varphi})\Big(K_{\chi}-2G_{\varphi}+G_{\chi\varphi}\dot{\varphi}^{2}\Big)+\dot{\varphi}^{2}\Big(K_{\chi\varphi}-G_{\varphi\varphi}\Big)+K_{\chi\chi}\dot{\varphi}^{2}\ddot{\varphi}-K_{\varphi}+\hspace{1.5cm}\nonumber\\+3G_{\chi}\Big(\dot{H}\dot{\varphi}^{2}+3H^{2}\dot{\varphi}^{2}+2\ddot{\varphi}\dot{\varphi}H\Big)-2G_{\chi\varphi}\dot{\varphi}^{2}\ddot{\varphi}+3HG_{\chi\chi}\dot{\varphi}^{3}\ddot{\varphi}=0.\hspace{0.1cm}
\label{new}    
\end{eqnarray}
In these equations, the quantity $a$ is the scale factor and $H=\dot{a}/a$ depicts the  Hubble parameter. In the following,  we will consider that the dots correspond to the derivative with respect to the cosmic time $t$ and  the subscribes "$\chi$", "$\chi\chi$",  "$\varphi$","$\varphi\chi$", etc, represent $\partial/\partial\chi$, $\partial^{2}/\partial\chi^{2}$  $\partial/\partial\varphi$, $\partial^{2}/\partial\varphi\partial\chi$, etc. %By varying the action (\ref{2}) with respect to the scalar field, the modified Klein-Gordon equation takes the following from
%\begin{eqnarray}
%3\dot{H}G_{\chi}\dot{\varphi}^{2}+\ddot{\varphi}\bigg(3HG_{\chi\chi}\dot{\varphi}^{3}-\dot{\varphi}^{2}\big(G_{\varphi\chi}-K_{\chi\chi}\big)+6HG_{\chi}\dot{\varphi}-2G_{\varphi}+K_{\chi}\bigg)+\hspace{1.5cm}\nonumber\\
%+3HG_{\varphi\chi}\dot{\varphi}^{3}+\dot{\varphi}^{2}\big(9H^{2}G_{\chi}-G_{\varphi\varphi}+K_{\varphi\chi}\big)-K_{\varphi}-3H\dot{\varphi}\big(2G_{\varphi}-K_{\chi}\big)=0.\hspace{0.1cm}
%\label{5}    
%\end{eqnarray}
In particular, for the special  case in which $K=\chi-V(\varphi)$ and $G=0$, where $V(\varphi)$ corresponds to   the effective potential associated to the  scalar field,  the expressions are reduced to standard  General Relativity (GR) with a single field. 

In order to fulfill the aim of the paper, we will analyze the G-inflation for the special case in which the function $K(\varphi,\chi)$ and the coupling function $G(\varphi,\chi)$ are defined as \cite{Herrera:2018wan,g0} 
\begin{equation}
K(\varphi, \chi)=\chi-V(\varphi),\hspace{1cm}G(\varphi,\chi)=g(\varphi)\chi^{n},
\label{6}
\end{equation}
respectively. However, we will also study a specific case in which  the coupling function $G(\varphi,\chi)$ is given by $G(\varphi,\chi)=\gamma_1+\gamma_2\varphi+\gamma_3\sqrt{\chi}$, with $\gamma_1$, $\gamma_2$  and $\gamma_3$ are constants.  

In the particular case $G(\varphi,\chi)=g(\varphi)\chi^{n}$, we consider $n>0$ and $g(\varphi)$ corresponds to a  coupling function depending only on the scalar field $\varphi$. Additionally, we define the function $g(\varphi)$ in terms of the scalar field as  $g(\varphi)=\gamma\varphi^{\nu}$ (power-law dependence) in which the parameters $\gamma>0$ and $\nu$ are real parameters \cite{Herrera:2018wan}. In this sense, the coupling parameter $G(\varphi,\chi)$ associated to the Galilean term $G(\varphi,\chi)\Box\varphi$, can be written as $G(\varphi,\chi)=\gamma\varphi^\nu\,\chi^n$. A particular case with $\nu=0$, \textit{i.e.}, $g(\varphi)=$ constant and $n\neq 0$  in the context of G-intermediate inflation was studied in Ref. \cite{g0}. Also, a special case with parameters $\nu$ and $n$ are different to zero in the framework of  the accelerated comic  expansions; intermediate, logamediate and exponential models was analyzed in \cite{Herrera:2018wan}.
 
In the context of the G-inflation, we will assume the slow-roll 
 approximation, in order to analyze the slow-roll equations together with the cosmological perturbations. Thus,
 under the slow-roll approximation, we can define the slow-roll parameters of the G-inflationary model as \cite{Kamada:2010qe}
\begin{equation}
\delta_{\chi}=\frac{K_{\chi}\chi}{H^{2}},\hspace{0.3cm}\delta_{G\chi}=\frac{G_{\chi}\dot{\varphi}\chi}{H},\hspace{0.3cm}\delta_{G\varphi}=\frac{G_{\varphi}\chi}{H^{2}},\hspace{0.3cm}\epsilon_{1}=-\frac{\dot{H}}{H^{2}},\hspace{0.3cm}\epsilon_{2}=-\frac{\ddot{\varphi}}{H\dot{\varphi}}=-\delta_{\varphi},\hspace{0.3cm}\epsilon_{3}=\frac{g_{\varphi}\dot{\varphi}}{gH},\hspace{0.3cm}\epsilon_{4}=\frac{g_{\varphi\varphi}\chi^{n+1}}{V_{\varphi}}.
\label{7}
\end{equation}
We mention that combined the above parameters and the dynamical equations (\ref{3}) and (\ref{4}), the first slow-roll parameter $\epsilon_{1}$ can be rewritten as
$\epsilon_{1}=\delta_{\chi}+3\delta_{G\chi}-2\delta_{G\varphi}-\delta_{\varphi}\delta_{G\chi}$. Additionally, we can introduce  another  slow-roll parameters defined as
\begin{equation}
\delta_{G\varphi\chi}=G_{\varphi\chi}\chi^{2}/H^{2},\hspace{0.5cm}\delta_{G\varphi\varphi}=G_{\varphi\varphi}\dot{\varphi}\chi/H^{3},\hspace{0.5cm}\delta_{G\varphi\chi\chi}=G_{\varphi\chi\chi}\chi^{3}/H^{2},\,\,\,\,\,\delta_{\chi\chi}=\frac{K_{\chi\chi}\chi^{2}}{H^{2}},\hspace{0.5cm}\delta_{G\chi\chi}=\frac{G_{\chi\chi}\dot{\varphi}\chi^{2}}{H}.
\label{9}
\end{equation}
In this form, considering the functions (\ref{6}) together with  the slow roll parameters defined by Eqs. (\ref{7}) and (\ref{9}), the equation of the motion for the scalar field (\ref{new})  applying the slow-roll conditions \textit{i.e.} $|\epsilon_{1}|$, $|\epsilon_{2}|$, $|\epsilon_{3}|$, $|\epsilon_{4}|\ll1$, is reduced to \cite{Kamada:2010qe,g2}
\begin{equation}
3H\dot{\varphi}(1+\mathcal{A})\simeq-V_{\varphi},
\label{11}
\end{equation}
%\textcolor{blue}{
%\begin{equation}
%3H\dot{\varphi}\mathcal{A}\simeq-V_{\varphi},
%\nonumber
%\end{equation}
%}
where the function $\mathcal{A}$ is defined as
\begin{equation}
\mathcal{A}\equiv3H\dot{\varphi}G_{\chi}.
%=3ng(\varphi)\chi^{n-%1}H\dot{\varphi}=3\gamma\,n\varp%hi^{\nu}\chi^{n-%1}H\dot{\varphi}.
\label{12}
\end{equation}
Interestingly, from the dynamics of the  Eq.(\ref{11}), we discriminate two different limits considering the behavior of the function $\mathcal{A}$ \text{e.g.} $|\mathcal{A}|\ll 1$ in which the standard slow roll equation for the scalar field coinciding GR is fulfilled.  In the opposite limit $|\mathcal{A}|\gg 1$, the Galileon term dominates the slow roll equation for the scalar field and then the Galilean effect alters the inflationary scenario.

On the other hand, for the sake of completeness, we briefly review the dynamical expressions of cosmological perturbations of G-inflation discussed in \cite{DeFelice:2013ar,g1,g2}. In this regard, the power spectrum of the scalar perturbation $\mathcal{P}_{\mathcal{S}}$ under the slow roll approximation is given by \cite{DeFelice:2013ar,g1,g2}
\begin{equation}
\mathcal{P}_{\mathcal{S}}\simeq\frac{H^{2}}{8\pi^{2}\varepsilon_{s}c_s},
\label{13}
\end{equation}
with the parameters $\varepsilon_{s}$ and $c_s$ are given by
\begin{equation}
\varepsilon_{s}=\delta_{\chi}+4\delta_{G\chi}-2\delta_{G\varphi},\,\,\,\,\,\mbox{and}\hspace{1cm}
c_s^2=\frac{3(2w_1^2w_2H-w_2^2-2w_1^2\dot{w}_2)}{w_1(4w_1w_3+9w_2^2)}
\label{14}
\end{equation}
respectively. Here the quantities $w_1$, $w_2$ and $w_3$ are defined as
$$
w_1=1,\,\,\,\,w_2=2H-2\chi\dot{\varphi}G_\chi,\,\,\,\,\mbox{and}\,\,\,\,w_3=-9H^2+3\chi+18H\dot{\varphi}(2\chi G_\chi+\chi^2G_{\chi\chi})-G\,\chi(G_\varphi+\chi G_{\varphi\chi}).
$$

For 
the scalar spectral index $n_s$ associated to the scalar power spectrum in the context of the slow roll approximation is defined as \cite{DeFelice:2013ar}
\begin{equation}
n_s-1=\frac{d\ln{\mathcal{P}_{S}}}{d\ln{k}}=-2\varepsilon_s-\eta_s-s,\label{ns2}
\end{equation}
where the parameters  $\eta_s$ and $s$ are given by
$$
\eta_s=\frac{\dot{\varepsilon}_s}{H\varepsilon_s},\,\,\,\,\,\mbox{and}\,\,\,\,\,s=\frac{\dot{c}_s}{Hc_s}.
$$

%where
%\begin{equation}
%\delta_{\chi\chi}=\frac{K_{\chi\chi}\chi^{2}}{H^{2}},\hspace{1cm}\del%ta_{G\chi\chi}=\frac{G_{\chi\chi}\dot{\varphi}\chi^{2}}{H}.
%\label{15}    
%\end{equation}
In particular for  the functions $K(\varphi,\chi)$ and $G(\varphi,\chi)$ given by Eq.(\ref{6}) and using the different slow-roll parameters,  the power spectrum (\ref{13}) under the slow-roll approximation is driven as \cite{g1,g2}
\begin{equation}
\mathcal{P}_{\mathcal{S}}\simeq\frac{H^{4}\sqrt{1+2n\mathcal{A}}}{8\pi^{2}\chi\sqrt{(1+\frac{4\mathcal{A}}{3})^{3}}}\simeq\frac{V^{3}(1+\mathcal{A})^{2}\sqrt{1+2n\mathcal{A}}}{12\pi^{2}V_{\varphi}^{2}\sqrt{(1+\frac{4\mathcal{A}}{3})^{3}}}.
\label{17}
\end{equation}
Here, we observe that in the specific limit  in which the function $|\mathcal{A}|\gg1$, where the effects of Galilean term dominates the inflationary stage, the expression for the power spectrum given by Eq.(\ref{17}) takes the form
%\begin{equation}
$
\mathcal{P}_{\mathcal{S}}\simeq\frac{3H^{4}\sqrt{6n}}{64\pi^{2}\chi\mathcal{A}}\simeq\frac{\sqrt{6n}V^{3}\mathcal{A}}{32\pi^{2}V_{\varphi}^{2}}.
$
%\label{18}
%\end{equation}
%\textcolor{blue}{
%\begin{equation}
%\mathcal{P}_{\mathcal{S}}\simeq3\sqrt{3}\frac{H^{4}}{64\pi^{2}}\frac{1}{\mathcal{A}\chi}.
%\nonumber
%\end{equation}
%}
In the context of G-inflation and considering Eq. (\ref{17}), the spectral index assuming the slow roll approximation results \cite{g1,g2}
\begin{equation}
n_{s}\simeq1-\frac{6\epsilon}{1+\mathcal{A}}+\frac{2\eta}{1+\mathcal{A}}+\frac{\dot{\mathcal{A}}}{H}\bigg(\frac{2}{1+\mathcal{A}}+\frac{n}{1+2n\mathcal{A}}-\frac{2}{1+\frac{4\mathcal{A}}{3}}\bigg),\
\label{19}
\end{equation}
where the slow-roll parameters $\epsilon$ and $\eta$ are defined in terms of the effective potential as
\begin{equation}
\epsilon=\frac{1}{2}\bigg(\frac{V_{\varphi}}{V}\bigg)^{2},\hspace{1cm}\eta=\frac{V_{\varphi\varphi}}{V}.
\label{20}
\end{equation}
From Eq.(\ref{19}), we note that in the specific case in which the function $\mathcal{A}\to 0$, the spectral index $n_s$ agrees with the expression found in the framework of the GR, in which $n_{s}-1\simeq-6\epsilon+2\eta$. Additionally, in the situation in which the  Galilean term dominates the inflationary scenario i.e.,
when the function $|\mathcal{A}|\gg 1$, the scalar spectral index is reduced to the expression 
%\begin{equation}
$n_{s}\simeq1-\frac{6\epsilon}{\mathcal{A}}+\frac{2\eta}{\mathcal{A}}+\frac{\dot{\mathcal{A}}}{H\mathcal{A}}.$
%\label{21}
%\end{equation}

In the context of the tensor perturbation, we mention that in the frame of G-inflation the equation of motion for the tensor perturbations remains the same as the framework of the RG,  see Refs. \cite{g1,g2}. In this way, the power spectrum associated with the primordial tensor modes can be written as  $\mathcal{P}_{G}=2H^{2}/\pi^{2}$. Thus, we can introduce the observational parameter $r$ defined as the tensor-to-scalar ratio $r=\mathcal{P}_{G}/\mathcal{P}_{\mathcal{S}}$ and in the context of  the slow-roll approximation we have
\begin{equation}
r=16\varepsilon_c\,c_s.
\end{equation}
As before, in the particular situation in which the coupling function is $G\propto \varphi^\nu\,\chi^n$, we have that the tensor to scalar ratio is reduced to
\begin{equation}
r\simeq16\epsilon\bigg(\frac{\sqrt{(1+\frac{4\mathcal{A}}{3})^{3}}}{(1+\mathcal{A})^{2}\sqrt{1+2n\mathcal{A}}}\bigg).
\label{19b}
\end{equation}
Here, in the specific situation in which the parameter 
 $\mathcal{A}\rightarrow0$, we can recover the expression for the  tensor-to-scalar ratio $r$ in the frame of the GR, where the parameter  $r$ is reduced to $r\simeq16\epsilon$. Also, in the limit in which the function  $|\mathcal{A}|\gg1$, the Eq.(\ref{19b}) associated to the tensor-to scalar ratio for G-inflation becomes
$r\simeq4\sqrt{\frac{2}{27}}\frac{16\epsilon}{\mathcal{A}\sqrt{n}}$, with $n\neq0$.
In the following, we will analyze the constant roll condition in the framework of G-inflation. In order to obtain analytical solutions for our model, we will study the simplest case in which the coupling function $G(\varphi,\chi)\propto\varphi$ \textit{i.e.}, $\nu=1$ and $n=0$. Also, in this scenario we will analyze the coupling function $G(\varphi,\chi)$ given by $G(\varphi,\chi)\propto\chi$, in which the powers $\nu=0$ and $n=1$, respectively. In both cases, we will study two specific situations  depending on the sign of the  arbitrary integration constants  associated to Hubble parameter. As a more rigorous case, we will study the model for the coupling function $G(\varphi,\chi)$ depending to both variables $\varphi$ and $\chi$. 

\section{G-constant-roll inflation}\label{GCR}

In this section, we will attempt to reconstruct the background variables (effective potential and Hubble parameter) together with the analysis of the cosmological perturbations in the framework of the G-inflation model with a constant rate of rolling. Starting from Eq. (\ref{4}) and using the constant-roll condition (\ref{1}) and the definition $\dot{H}=\dot{\varphi}H_{\varphi}$, we have 
\begin{equation}
2\dot{\varphi}H_{\varphi}+3H^{2}+\chi-V(\varphi)-\dot{\varphi}^{2}\Big(G_{\varphi}-G_{\chi}(\alpha+3)H\dot{\varphi}\Big)=0.
\label{n1}    
\end{equation}
Then, considering the Friedmann equation (\ref{3}), the above expression can be rewritten  as
\begin{equation}
\dot{\varphi}\bigg(2H_{\varphi}+H\dot{\varphi}^{2}G_{\chi}(\alpha+6)+\dot{\varphi}(1-2G_{\varphi})\bigg)=0.
\label{n2}    
\end{equation}
Here, we have assumed $K(\varphi, \chi)=\chi-V(\varphi)$. In the following, we will  study the constant-roll inflation for some specific forms of the coupling function $G(\varphi,\chi)\propto\varphi^\nu\,\chi^n$ or equivalently $\nu$ and $n$, when the speed of the scalar field $\dot{\varphi}\neq0$. Also, a specific form of the function $G$ as a mixture of the variables $\varphi$ and $\chi$ will be investigated.

\subsection{CASE I}

In order to find an analytical solution, we study the simplest case in which the powers $\nu=1$ and $n=0$, with the function $G(\varphi,\chi)$ 
\begin{equation}
G(\varphi,\chi)=\gamma \varphi,
\label{n3}    
\end{equation}
where $\gamma$ is a constant parameter. Then, from Eq. (\ref{n2}), we find the  expression for $\dot{\varphi}$ as follows 
\begin{equation}
\dot{\varphi}=\frac{-2H_{\varphi}}{(1-2\gamma)},\,\,\,\,\,\mbox{with}
\,\,\,\,\,\gamma\neq\frac{1}{2}.
\label{n4}   
\end{equation}
Now, taking the  derivative of Eq.(\ref{n4}) with respect to the cosmic time and  considering  the constant-roll condition (\ref{1}), we obtain that the differential equation for the Hubble parameter $H$ as a function of the scalar field $\varphi$ results 
\begin{equation}
H_{\varphi\varphi}=\frac{(3+\alpha)(1-2\gamma)}{2}H.
\label{n5}    
\end{equation}
We note that by setting $\gamma=0$ (or $G=0$), we recover the differential equation of $H$ for the single field constant-roll inflation introduced in \cite{Motohashi1}. The general solution of the above differential equation can be written as
\begin{equation}
H(\varphi)=C_{1}\exp\bigg(\sqrt{\frac{(3+\alpha)(1-2\gamma)}{2}}\varphi\bigg)+C_{2}\exp\bigg(-\sqrt{\frac{(3+\alpha)(1-2\gamma)}{2}}\varphi\bigg),
\label{n6}    
\end{equation}
where $C_1$ and $C_2$ correspond to two arbitrary integration constants. We observe that for values of $\alpha<-3$ and $\gamma<1/2$, we encounter with an imaginary form of the exponential function (or a simple harmonic oscillator motion equation) of the parameter $H(\varphi)$. Similarly, we have one harmonic oscillator equation, if we consider the values of $\alpha>-3$ and $\gamma>1/2$, respectively. From Eq. (\ref{n6}), interesting solutions occur when the values of the parameters $\alpha$ and $\gamma$ are $\alpha>-3$ and $\gamma<1/2$ (or also $\alpha<-3$ and $\gamma>1/2$). In this respect, we mention that unlike the standard constant-roll inflation, we have the possibility to consider the values in which $\alpha<-3$ (with $\gamma>1/2$).  
In the following, we will discuss these solutions depending on the values of the integration constants $C_1$ and $C_2$. In this form, by assuming $C_1\,C_2=0$, from Eq. (\ref{n6}), we have the solution for the Hubble parameter as $H(\varphi)\propto\exp\big(\pm\sqrt{\frac{(3+\alpha)(1-2\gamma)}{2}}\varphi\big)$. Also, as another interesting solution, we have the situation in which the constant $C_{1}=C_{2}=M/2$, where the quantity $M>0$ corresponds to a constant  associated to energy scale. Therefore, the Hubble parameter can be rewritten as
\begin{equation}
H(\varphi)=M\cosh{\bigg(\sqrt{\frac{(3+\alpha)(1-2\gamma)}{2}}\varphi\bigg)}.
\label{n7}    
\end{equation}
%where $M$ is an integration constant which depicts the %amplitude of the power spectrum of the curvature %perturbation.
By using Eq. (\ref{n4}) and the Friedmann equation (\ref{3}), the speed $\dot{\varphi}$ and the effective potential associated to the scalar field $\varphi$  are given by
\begin{equation}
\dot{\varphi}=-M\sqrt{\frac{2(3+\alpha)}{1-2\gamma}}\sinh{\bigg(\sqrt{\frac{(3+\alpha)(1-2\gamma)}{2}}\varphi\bigg)},
\label{n8}    
\end{equation}
and
\begin{equation}
V(\varphi)=3M^{2}\bigg(1+\frac{\alpha}{6}\Big[1-\cosh{\big(\sqrt{2(3+\alpha)(1-2\gamma)}\varphi\big)}\Big]\bigg).
\label{n9}    
\end{equation}
Also, the scale factor $a(t)$ of the model takes the following form
\begin{equation}
a(t)\propto\sinh^{\frac{1}{3+\alpha}}\Big((3+\alpha)Mt\Big),
\label{n10}
\end{equation}
and the scalar field in terms of the cosmic time $t$ results
\begin{equation}
\varphi(t)=\sqrt{\frac{2}{(3+\alpha)(1-2\gamma)}}\ln{\bigg(\coth{\Big(\frac{M(3+\alpha)t}{2}\Big)}\bigg)}+C_0,
\label{rh1}
\end{equation}
where $C_0$ is an integration constant.

To complete the analysis of our constant-roll inflationary model with the  coupling function $G(\varphi,\chi)\propto \varphi$, in the following we will analyze the effective  potential given by Eq.(\ref{n9}) in the case in which  $\alpha<-3$ and $\gamma<1/2$ (or also $\alpha>-3$ and $\gamma>1/2$). For the specific case $\alpha<-3$ and $\gamma<1/2$, the reconstructed potential can be written as

\begin{equation}
V(\varphi)=3M^{2}\bigg(1-\frac{3+\beta}{6}\Big[1-\cos{\big(\sqrt{2\beta(1-2\gamma)}\varphi\big)}\Big]\bigg),
\label{n11}    
\end{equation}
where the parameter $\beta $ is defined as
$\beta=-(3+\alpha)>0$. 

Thus, for the special case when $\beta>0$ (or equivalently $\alpha<-3$) together $\gamma<1/2$ and from Eq.(\ref{n7}), we find the Hubble parameter as $H(\varphi)=M\,\cos(\sqrt{|(3+\alpha)|(1-2\gamma)/2}\,\,\varphi)$. Then, considering Eq.(\ref{n4}), we obtain the velocity of the scalar field as
\begin{equation}
\dot{\varphi}=M\sqrt{\frac{2|(3+\alpha)|}{1-2\gamma}}\sin{\bigg(\sqrt{\frac{|(3+\alpha)|(1-2\gamma)}{2}}\varphi\bigg)}.
\label{n8b}    
\end{equation}
In this way, the solution of the scalar field as a function of the time can be found as
\begin{equation}
\varphi(t)=\frac{2\sqrt{2}}{\sqrt{|(3+\alpha)|(1-2\gamma)}}
\arctan\left[ e^{M|(3+\alpha)|\,t+C_0}\right],
\label{soln8b}    
\end{equation}
where $C_0=\sqrt{|(3+\alpha)|(1-2\gamma)}\,\,C_1$ with
$C_1$  an integration constant.

Now, if the parameters $\alpha>-3$ and $\gamma>1/2$, from Eq.(\ref{n9}), the potential of the scalar field take the form
\begin{equation}
V(\varphi)=3M^{2}\bigg(1+\frac{\alpha}{6}\Big[1-\cos{\big(\sqrt{2(3+\alpha)\,\tilde{\gamma}}\,\varphi\big)}\Big]\bigg),\hspace{1cm}\mbox{where}\hspace{1cm}\tilde{\gamma}=|1-2\gamma|.
\label{n11b}    
\end{equation}
In the following and for simplicity, we consider the case 
$\beta>0$ and $\gamma<1/2$. Thus, we can calculate the slow-roll parameters (\ref{20}) of the model as
\begin{equation}
\epsilon=\frac{\beta(3+\beta)^{2}(1-2\gamma)\sin^{2}\Big(\sqrt{2\beta(1-2\gamma)}\varphi\Big)}{\Big(6-(3+\beta)\big(1-\cos\big(\sqrt{2\beta(1-2\gamma)}\varphi\big)
\big)\Big)^{2}},
\label{n12}    
\end{equation}
and
\begin{equation}
\eta=-\frac{2\beta(3+\beta)(1-2\gamma)\cos\Big(\sqrt{2\beta(1-2\gamma)}\varphi\Big)}{6-(3+\beta)\big(1-\cos\big(\sqrt{2\beta(1-2\gamma)}\varphi\big)\big)}.   
\label{n13}    
\end{equation}
During the inflationary scenario, the slow roll parameters $\epsilon$ and $\eta$ must be sufficiently small and this scenario  ends when $\epsilon=1$ or $\eta=1$ is fulfilled. Also, we can introduce  the number of $e$-folds $N$ 
\begin{figure*}[!hbtp]
	\centering
    \includegraphics[width=.40\textwidth,keepaspectratio]{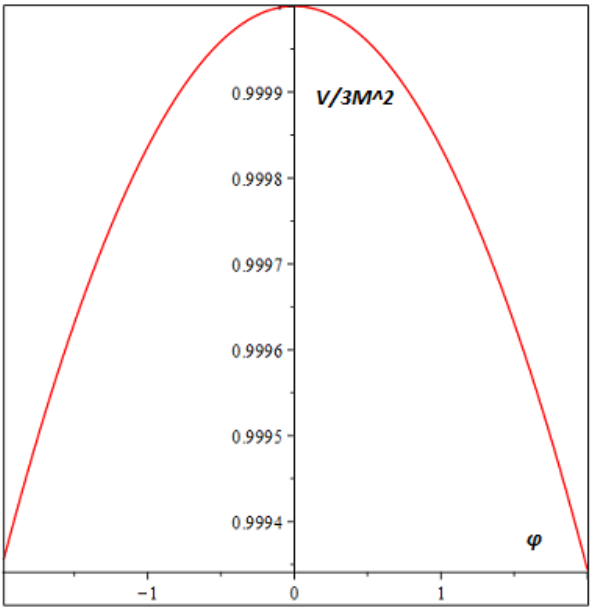}
	\hspace{1cm}
	\includegraphics[width=.40
	\textwidth,keepaspectratio]{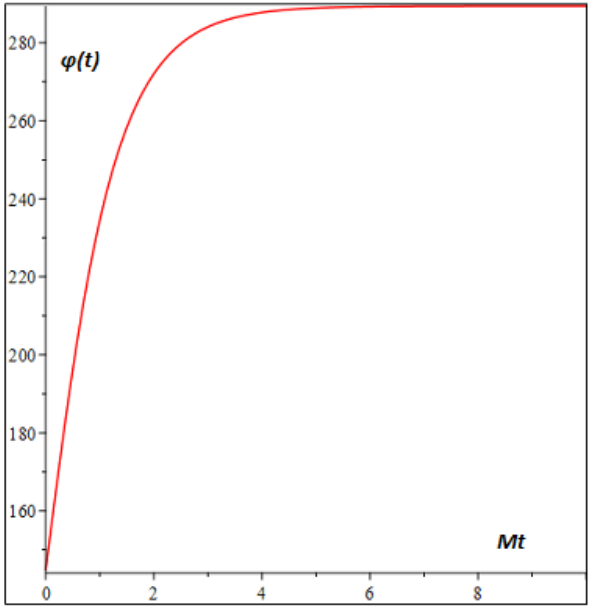}
	\caption{The effective potential versus the scalar field (\ref{n11}) and the evolution of inflaton versus cosmic time $t$ (\ref{soln8b}). The plots are drawn for the obtained values of the model \textit{i.e.} $\beta=1.703>0$ (or equivalently $\alpha=-4.703$) and the parameter $\gamma=0.4990$.}
	\label{fig1}
\end{figure*}
\begin{equation}
N=\int_{t_*}^{t_f}H\, dt\simeq\int^{\varphi_{*}}_{\varphi_{f}}{\frac{1}{\sqrt{2\epsilon}}}d\varphi\simeq\frac{\ln\bigg(\Big(\cos\big(\sqrt{2\beta(1-2\gamma)}\varphi_{*}+1\big)\Big)^{2\beta}\Big(\cos\big(1-\sqrt{2\beta(1-2\gamma)}\varphi_{*}\big)\Big)^{6}\bigg)}{4\beta(3+\beta)(1-2\gamma)},
\label{n14}
\end{equation}
which is defined between two values of cosmological times $t_*$ and at the end of inflation $t_f$ or equivalently between two different values of the scalar field 
 $\varphi_*$ and $\varphi_f$.  In what follows the subscript $*$ is utilized to indicate the epoch where the cosmological scale exists the horizon. 

Now, from Eq.(\ref{19}) and 
using the slow-roll parameters (\ref{n12}) and (\ref{n13}) and the above expression for the number of e-folds $N$, we find that the curvature power spectrum (\ref{17}), 
%(in the case $\mathcal{A}\rightarrow 0$),
the scalar spectral index and the tensor-to-scalar ratio in terms of $N$ become
\begin{equation}
\mathcal{P_{\mathcal{S}}}=\frac{\big(3+\beta-6\mathcal{Q}^{-1}\big)^{3}M^{2}\mathcal{Q}}{48\big(2\mathcal{Q}^{-1}-1\big)\beta\pi^{2}(1-2\gamma)(3+\beta)^{2}},
\label{PR1}    
\end{equation}
\begin{equation}
n_{s}\simeq \frac{4\Big(-3+2(2\gamma-1)\beta^{2}\Big)(3+\beta)\mathcal{Q}-\Big(-1+2(2\gamma-1)\beta\Big)(3+\beta)^{2}\mathcal{Q}^{2}+36+24(2\gamma-1)\beta^{2}+72(2\gamma-1)\beta}{\Big(6-(3+\beta)\mathcal{Q}\Big)^{2}},
\label{n15}    
\end{equation}
and 
\begin{equation}
r\simeq \frac{16\,\beta(3+\beta)^{2}(2\gamma-1)\mathcal{Q}(\mathcal{Q}-2)}{\Big(6-(3+\beta)\mathcal{Q}\Big)^{2}},\label{r1}
\end {equation}
where $\mathcal{Q}(N)=\mathcal{Q}=\exp{(-\frac{2N\beta(3+\beta)(2\gamma-1)}{3})}$.
Here we observe that both the index $n_s(N)$ and the ratio $r(N)$, are independent of the parameter associated to the energy scale $M$. 

We note that from Eq.(\ref{PR1}), we can obtain an expression for the energy scale $M>0$, in terms of the number of $e$-folds $N$ and the scalar power spectrum $\mathcal{P_{\mathcal{S}}}$ result 
\begin{equation}
M=\sqrt{\frac{48\pi^{2}(3+\beta)^{2}\beta(1-2\gamma)(-1+2\mathcal{Q}^{-1})\mathcal{P_{\mathcal{S}}}}{(\beta+3-6\mathcal{Q}^{-1})^{3}\mathcal{Q}}}.
\label{M1}    
\end{equation}
\begin{figure*}[!hbtp]
	\centering
    \includegraphics[width=.40\textwidth,keepaspectratio]{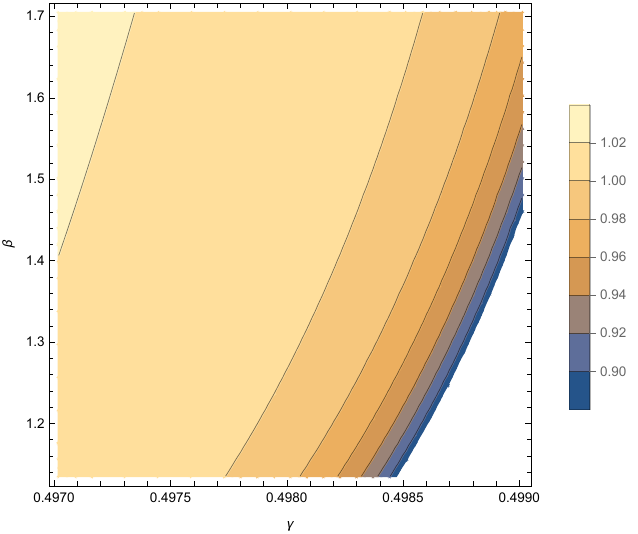}
	\hspace{1cm}
	\includegraphics[width=.40
	\textwidth,keepaspectratio]{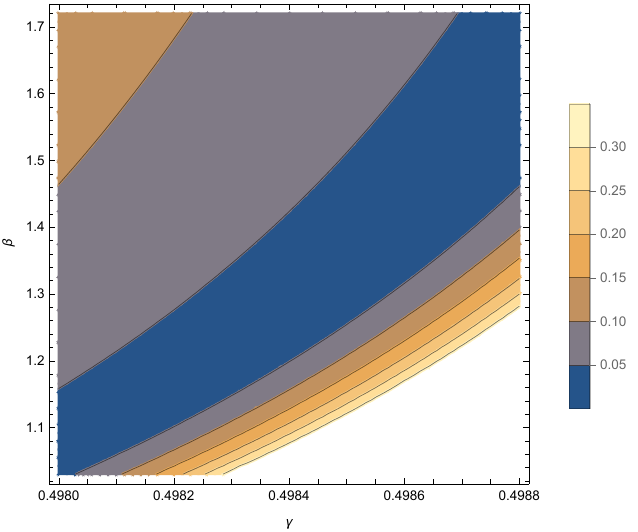}
	\caption{The left panel corresponds to the contour plot for the scalar spectral index $n_s$ as a function of the parameters $\beta$ and $\gamma$ given by Eq.(\ref{n15}). The  right panel represents  the contour plot for the tensor to scalar ratio  $r$ as a function of the parameters $\beta$ and $\gamma$ from Eq.(\ref{r1}). In both panels we have fixed the number of $e-$folds $N=60$.   }
	\label{fig1AA}
\end{figure*}
\begin{figure*}[!hbtp]
	\centering
	\includegraphics[width=.70\textwidth,keepaspectratio]{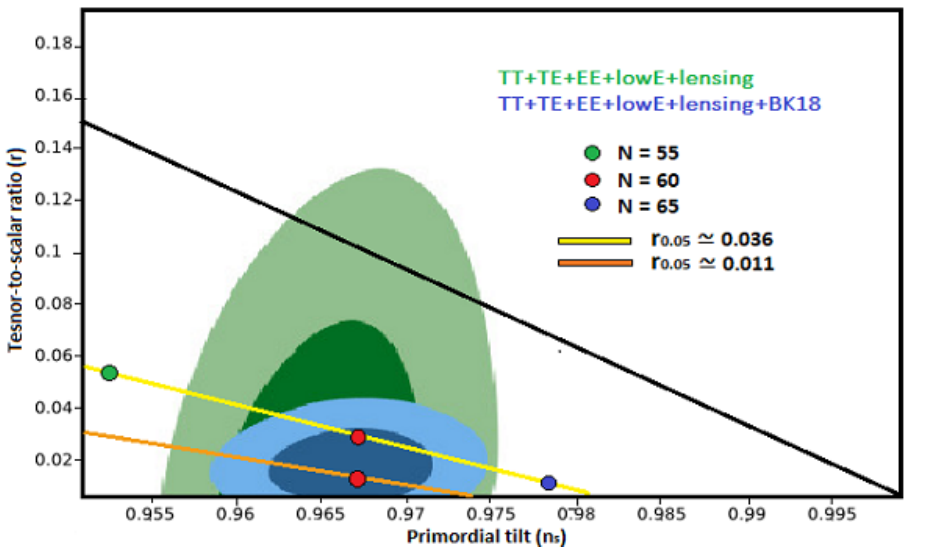}
	\caption{The marginalized joint 68\% and 95\% CL regions for $n_{s}$ and $r$ plane   from Planck alone and in combination with  BK18+BAO data \cite{BICEP:2021xfz} and the constraints in the $r-n_{s}$ on the model (\ref{n3}). The BK18+BAO data places a stronger limits on the tensor to scalar ratio $r$ (shown in blue). The blue contours correspond to the constraint on the tensor to scalar ratio tightens from $r_{0.05}<0.11$ to $r_{0.05}<0.035$. In this panel, the lines with orange and yellow colors correspond to the pairs ($\beta=1.202$, $\gamma=0.498$) and  ($\beta=1.703$, $\gamma=0.4990$), when the energy scale  $M=1.767\times 10^{-5}$ and $M=2.892\times 10^{-5}$ respectively.}
	\label{fig1a}
\end{figure*}

Additionally, we can find the constraints for the parameter $\gamma$ associated to the function $G(\varphi, \chi)$ together with the parameter $\beta=-(3+\alpha)$ and the quantity $M$, when the number of $e-$folds $N$, the scalar power spectrum $\mathcal{P_{\mathcal{S}}}$, the scalar spectral index $n_s$ and the tensor to scalar ratio $r$ are  given. In this context, considering  the observational parameters   $n_s=0.967$ and  $r=0.036$ from Plank data \cite{BICEP:2021xfz} at $N=60$ and from Eqs.(\ref{n15}) and (\ref{r1}), we find the constraint for $\beta$ as $\beta=1.703>0$ (or equivalently $\alpha=-4.703$) and for $\gamma$ as $\gamma=0.4990<1/2$. Moreover, assuming that the power spectrum $\mathcal{P_{\mathcal{S}}}=2.2\times 10^{-9}$ and from Eq.(\ref{M1}), we obtain the constraint for $M$ as $M=2.892\times 10^{-5}$. Also, for the case $r=0.011$ and $n_s=0.967$ evaluated at $N=60$, we get the parameters $\beta=1.202$ and $\gamma=0.498$, numerically. Thus, from Eq.(\ref{M1}), the energy scale becomes $M=1.767\times 10^{-5}$.

Now considering the lower bound for the tensor to scalar ratio in which $r=0$ together with $n_s=0.967$ at the value $N=60$, we find the numerical solution $\beta=1.051$, $\gamma=0.497$ and the energy scale $M=1.566\times 10^{-11}$. In this case, we find a narrow range for the parameters $\gamma$ and $\beta$. While the energy scale $M$ decreases about six orders of magnitude so that the range for the energy scale in order to satisfy the observational constraints corresponds to $10^{-11}<M<10^{-5}$.

The left panel of Fig.\ref{fig1} shows the behaviour of the effective potential given by Eq.(\ref{n9}) as a function of the scalar field $\varphi$, for the values of the parameters $\beta=1.703$ and  $\gamma=0.4991$ according to the observational parameters. 
%In particular we also consider  the case of $\alpha=0$ that depicts the ultra slow-roll %version of the model \cite{g3}, in which the effective potential becomes constant i.e., %$V(\varphi)=3M^2=$ constant. 
In the right panel, we present the evolution of inflaton in term of the cosmic time given by Eq.(\ref{rh1}), for the values of $\alpha$ and $\gamma$ found from the observational parameter and additionally we have considered that the integration constant  $C_{0}=0$.

In Fig.\ref{fig1AA} the left panel shows the contours curves associated with the same scalar spectral index $n_s$, and different combinations of the parameters $\beta$ (or analogously $\alpha$) and $\gamma$ by fitting Eq.(\ref{n15}). Similarly the right panel shows the contours curves associated with the same tensor to scalar ratio assuming different values of $\beta$ and $\gamma$ from Eq.(\ref{r1}). In both panels we have used the number of $e-$folds $N=60$. From the left panel, given a value of the scalar spectral index $n_s$ (vertical column), one can therefore constrain  the parameter space for the quantities  $\beta$ and $\gamma$.  The same way, from the right panel for the tensor to scalar ratio $r$, one can  constrain  the parameter space ($\beta$,$\gamma$). In both panels, we  can observe that narrow range for the parameter $\gamma$ in order to satisfy the observational constraint from the Planck data.

In Fig.\ref{fig1a} we show the consistency relation $r=r(n_{s}) $  coming from the marginalized joint 68\% and 95\% CL regions of the Planck 2018 alone and when adding in   BK18+BAO data \cite{BICEP:2021xfz} on the G-inflation model (\ref{n3}) in the constant-roll regime. In this figure, the lines with orange and yellow colors  correspond to the pairs  ($\beta=1.202$, $\gamma=0.498$) and  ($\beta=1.703$, $\gamma=0.4990$), respectively. The green contours of this figure is adjusted  of Ref.\cite{Planck:2018vyg}.

\subsection{CASE II}
As a second  case, we consider the powers $\nu$ and $n$ as $\nu=0$ and $n=1/2$ so that  
function $G(\varphi,\chi)$ from  Eq.(\ref{6}) becomes
\begin{equation}
G(\varphi,\chi)=g_{0}\sqrt{\chi}\,,
\label{k1}    
\end{equation}
where $g(\varphi)=g_{0}$ is a constant parameter with the dimension of $[mass^{-1}]$ \cite{g0}. From eq.(\ref{n2}) and using the definition $\dot{\varphi}=\sqrt{2\chi}$, we find $\dot{\varphi}$ as
\begin{equation}
\dot{\varphi}=-\frac{4H_{\varphi}}{2+Hg_{0}\sqrt{2}(6+\alpha)}.
\label{k2}    
\end{equation}
By taking time derivation of the above equation and then using the constant-roll condition (\ref{1}), we obtain
\begin{equation}
-(3+\alpha)H=-\frac{8H_{\varphi\varphi}+4g_{0}\sqrt{2}(6+\alpha)(H_{\varphi\varphi}H-H_{\varphi}^{2})}{\big(2+Hg_{0}\sqrt{2}(6+\alpha)\big)^{2}}.
\label{k3}    
\end{equation}
Obviously, in the limit of $g\rightarrow0$, we recover the standard constant-roll inflation for a single scalar field discussed in Ref.\cite{Motohashi1}. By introducing the variable $y=(1/A)\ln[2+AH]$ then:
\begin{equation}
  y_\varphi=\frac{H_\varphi}{2+A\,H},\,\,\mbox{with } \,\,\,A=g_0\sqrt{2}(6+\alpha)\,\,\,\mbox{and }\dot{\varphi}=-4y_\phi.
  \label{k4}
\end{equation}
Thus, Eq.(\ref{k3}) can be rewritten as
\begin{equation}
y_{\varphi\varphi}=\frac{(3+\alpha)}{4A}(e^{Ay}-2).
\label{k5}
\end{equation}
By assuming that during the evolution of the inflationary epoch the Hubble parameter $H\gg 2/A$, we have
\begin{equation}
 y_{\varphi\varphi}\simeq\frac{(3+\alpha)}{4A}e^{Ay}.
\label{k6}
\end{equation}
This equation can be solved as
\begin{equation}
y(\varphi)=\frac{1}{A}\ln\Big[-\frac{2A^2C_{1}}{3+\alpha}\frac{1}{\cosh^{2}\Big(\frac{A(\varphi+C_{2})\sqrt{C_{1}}}{2}\Big)}\Big],
\label{k7}
\end{equation}
where $C_1$ and $C_2$ correspond to two integration constants.   
Thus, from Eq.(\ref{k7}), we find that the Hubble parameter in terms of the scalar fields yields
\begin{equation}
H(\varphi)=-\frac{2\sqrt{2}g_{0}(6+\alpha)C_{1}}{(3+\alpha)}\frac{1}{\cosh^{2}\Big(\frac{g_{0}(6+\alpha)(\varphi+C_{2})\sqrt{C_{1}}}{\sqrt{2}}\Big)}.\label{k8}
\end{equation}
%where $A=g_{0}\sqrt{2}(6+a)$. 

Since the Hubble parameter is a positive quantity and assuming the parameter $g_0>0$, we include  a new  condition given by $(6+\alpha)C_1/(3+\alpha)<0$ that can be satisfied by two configurations as follow:
\begin{equation}
  H(\varphi)=\begin{cases}
    \frac{2\sqrt{2}g_{0}(6+\alpha)C_{1}}{|3+\alpha|}\frac{1}{\cosh^{2}\Big(\frac{g_{0}(6+\alpha)(\varphi+C_{2})\sqrt{C_{1}}}{\sqrt{2}}\Big)}, & \text{if $C_{1}>0$ and $\frac{(6+\alpha)}{(3+\alpha)}<0$ (or $-6<\alpha<-3$)}.\\
    \frac{2\sqrt{2}g_{0}(6+\alpha)|C_{1}|}{(3+\alpha)}\frac{1}{\cos^{2}\Big(\frac{g_{0}(6+\alpha)(\varphi+C_{2})\sqrt{|C_{1}|}}{\sqrt{2}}\Big)}, & \text{if $C_{1}<0$ and $\frac{(6+\alpha)}{(3+\alpha)}>0$\,(with which $\alpha<-6$ or $\alpha>-3$)}.
  \end{cases}
  \label{k9}
\end{equation}

On the other hand, from the Friedmann equation (\ref{3}) and considering the term $\dot{\varphi}$ (\ref{k2}) under the approximation $H\gg2/A$, we find that the effective potential results
\begin{equation}
V(\varphi)=3H^{2}-\frac{4(1+3Hg_0\sqrt{2})}{g_0^2(6+\alpha)^{2}}\frac{H_{\varphi}^{2}}{H^2}.
\label{k10}    
\end{equation}
In the following we will analyze the different solutions for the Hubble parameter given  by Eq.(\ref{k9}) from the condition that relates the integration constant $C_1$ and the $\alpha$ parameter through  $(6+\alpha)C_1/(3+\alpha)<0$. 

\subsubsection{Solution 1}

As the first case, we consider the solution of the Hubble parameter (\ref{k8}) with $C_1>0$ and $\frac{(6+\alpha)}{(3+\alpha)}<0$ (or equivalently $-6<\alpha<-3$) that leads to
\begin{equation}
H(\varphi)=\frac{2\sqrt{2}g_{0}(6+\alpha)C_{1}}{|3+\alpha|}\frac{1}{\cosh^{2}\Big(\frac{g_{0}(6+\alpha)(\varphi+C_{2})\sqrt{C_{1}}}{\sqrt{2}}\Big)}.
\label{k11}    
\end{equation}
Then, plugging the above expression in eq.(\ref{k10}), we find
\begin{eqnarray}
V(\varphi)=\frac{24C_{1}^2g_{0}^2(6+\alpha)}{|3+\alpha|\cosh^{4}\Big(\frac{g_{0}(6+\alpha)(\varphi+C_{2})\sqrt{C_{1}}}{\sqrt{2}}\Big)}\Bigg\{\Big(\frac{6+\alpha}{|3+\alpha|}+4\Big)-\Big(4-\frac{|3+\alpha|}{3g_{0}^2C_{1}(6+\alpha)}\Big)\times\hspace{3cm}\nonumber\\
\times\cosh^{2}\Big(\frac{g_{0}(6+\alpha)(\varphi+C_{2})\sqrt{C_{1}}}{\sqrt{2}}\Big)-\frac{|3+\alpha|}{3g_{0}^2C_{1}(6+\alpha)}\cosh^{4}\Big(\frac{g_{0}(6+\alpha)(\varphi+C_{2})\sqrt{C_{1}}}{\sqrt{2}}\Big)\Bigg\}.\hspace{0.5cm}
\label{k12}    
\end{eqnarray}
Also, the speed of the scalar field $\varphi$ (\ref{k2}) is given by
\begin{equation}
\dot{\varphi}=4\sqrt{C_{1}}\tanh\Big(\frac{g_{0}(6+\alpha)(\varphi+C_{2})\sqrt{C_{1}}}{\sqrt{2}}\Big).
\label{k13}
\end{equation}
Now, by introducing a new scalar field $\phi=\varphi+C_{2}$ and also choosing $C_{1}=M^2$, the Hubble parameter (\ref{k11}) can be found as
\begin{equation}
H(\phi)=\frac{H_0}{\cosh^2\Big(\frac{g_{0}(6+\alpha)}{\sqrt{2}}M\phi\Big)},\hspace{0.2cm}\text{where}\hspace{0.5cm}H_{0}=\frac{2\sqrt{2}g_{0}(6+\alpha)M^2}{|3+\alpha|}.
\label{k14}    
\end{equation}
Moreover, the reconstructed potential (\ref{k12}) is rewritten as
\begin{eqnarray}
V(\phi)=\frac{V_0}{\cosh^{4}\Big(\frac{g_{0}(6+\alpha)}{\sqrt{2}}M\phi\Big)}\Bigg\{\Big(\frac{6+\alpha}{|3+\alpha|}+4\Big)-\hspace{8cm}\nonumber\\
-\Big(4-\frac{|3+\alpha|}{3g_{0}^2M^2(6+\alpha)}\Big)\cosh^{2}\Big(\frac{g_{0}(6+\alpha)}{\sqrt{2}}M\phi\Big)-\frac{|3+\alpha|}{3g_{0}^2M^2(6+\alpha)}\cosh^{4}\Big(\frac{g_{0}(6+\alpha)}{\sqrt{2}}M\phi\Big)\Bigg\},\hspace{0.4cm}
\label{k15}    
\end{eqnarray}
where $V_{0}=\frac{24M^4g_{0}^2(6+\alpha)}{|3+\alpha|}$. From Eq.(\ref{k13}), we find the evolution of the new scalar field $\phi$ versus cosmic time $t$ as
\begin{equation}
\phi(t)=\phi_0\sinh^{-1}\Big(\exp\Big[\frac{4g_{0}(6+\alpha)}{\sqrt{2}}M^{2}t\Big]\Big),\,\,\,\,\,\,\text{where}\hspace{0.5cm}\phi_{0}=\frac{\sqrt{2}}{g_{0}(6+\alpha)M},
\label{k16}    
\end{equation}
and then, plugging the above relation into Eq.(\ref{k14}), the Hubble parameter $H$ and scale factor $a$ versus cosmic time are given by
\begin{equation}
H(t)=\frac{H_0}{1+\exp\Big[\frac{8g_{0}(6+\alpha)}{\sqrt{2}}M^{2}t\Big]},\hspace{1cm} \mbox{and}\,\,\,\,a(t)\propto\frac{e^{H_0t}}{\Big(1+e^{2|3+\alpha|H_0t}\Big)^{\frac{1}{2|3+\alpha|}}},
\label{k17}    
\end{equation}
 respectively. Notice that in the specific case in which $ e^{2|3+\alpha|H_0t}\sim 1$ (or equivalently to times $t\sim 0$), the scale factor corresponds to the exponential expansion or de Sitter-expansion, since $a(t)\propto e^{H_0t}$ and in the  limit, the scale factor becomes a constant. Since the scale factor (\ref{k17}) shows $\ddot{a}>0$ for early times, the solution (\ref{k11}) is an inflationary solution. 

In order to study the cosmological perturbations of the model, 
%in a viable inflationary framework, 
%for the rest of paper, 
we will work with the accredited potential (\ref{k15}) in a specific regime $(6-6\delta)/\delta<\alpha<-3$ (with $\delta=3g_{0}\sqrt{2}H>2$) in which the third term in the parenthesis is enough small to be negligible with respect to the second term. Therefore, the potential of the model in terms of the scalar field $\phi$ reduces to
%$\beta=-(3+\alpha)>0$ is rewritten as
\begin{equation}
V(\phi)\approx\frac{V_0}{\cosh^{4}\Big(\frac{g_{0}(3-\beta)}{\sqrt{2}}M\phi\Big)}\Bigg\{\frac{3(1+\beta)}{\beta}-4\cosh^{2}\Big(\frac{g_{0}(3-\beta)}{\sqrt{2}}M\phi\Big)\Bigg\},
\label{k28}    
\end{equation}
where $V_{0}=\frac{24M^4g_{0}^2(3-\beta)}{\beta}$ and the quantity $\beta$ is defined as $\beta=-(3+\alpha)$ and then the range for $\beta$ corresponds to $3>\beta>0$. In this context, we  find that the slow-roll parameters defined by Eq.(\ref{20}) become 
\begin{equation}
\epsilon=\frac{4g_{0}^{2}M^2(3-\beta)^2\tanh^{2}\Big(\frac{g_{0}(3-\beta)}{\sqrt{2}}M\phi\Big)\Big(2\beta\cosh^2\Big(\frac{g_{0}(3-\beta)}{\sqrt{2}}M\phi\Big)-3(\beta+1)\Big)^{2}}{\Big(4\beta\cosh^2\Big(\frac{g_{0}(3-\beta)}{\sqrt{2}}M\phi\Big)-3(\beta+1)\Big)^{2}},
\label{k29}
\end{equation}
and
\begin{equation}
\eta=\frac{2g_{0}^{2}M^2(3-\beta)^2\sech^{2}\Big(\frac{g_{0}(3-\beta)}{\sqrt{2}}M\phi\Big)\Big(4\beta\cosh^4\Big(\frac{g_{0}(3-\beta)}{\sqrt{2}}M\phi\Big)-6(3\beta+2)\cosh^2\Big(\frac{g_{0}(3-\beta)}{\sqrt{2}}M\phi\Big)+15(\beta+1)\Big)}{4\beta\cosh^2\Big(\frac{g_{0}(3-\beta)}{\sqrt{2}}M\phi\Big)-3(\beta+1)},
\label{k30}
\end{equation}
respectively. 
As during the constant-roll inflation $\epsilon\ll1$
%(while $\eta$ is not negligible) 
(or equivalently $\ddot{a}\gg$ 0) while 
%and 
inflation terminates when $\epsilon=1$ is satisfied. Thus, by setting $\epsilon=1$, we find that the real solution for the value at the end of inflation for the scalar field $\phi_f$ becomes 
\begin{equation}
\phi_f=\Big(\frac{\sqrt{2}}{g_0BM}\Big)\cosh^{-1}\left[\frac{4\beta^3(\beta+B(8A_1^2-1))+2\beta B(2A_1^2-1)L_1^{1/3}+L_1^{2/3}+L_2}{6(A_1^2-1)\beta^2\,L_1^{1/3}}\right]^{1/2},\label{New6}
\end{equation}
where the constants $L_1$ and $L_2$ are given by
\begin{eqnarray}
L_1=4\beta^{5}(3B-2\beta)-3\beta^4B^2(2-17A_1^2+68A_1^4)+\beta^3B^3(1-15A_1^2+30A_1^4-8A_1^6)\,+\hspace{3cm}
\nonumber\\
3\sqrt{3}\sqrt{\beta^6B^3(A_1-4A_1^3)\Big[8\beta^3+\beta^2B(71A_1^2-12)+2\beta B(3-13A_1^2+2A_1^4 )-(A_1^2-1)^2B^3\Big]},
\end{eqnarray}
and
\begin{equation}
L_2=\beta^2\Big[(1-A_1^2+4A_1^4)B^2-2L_1\Big]^{1/3},
\end{equation}
with the quantities $A_1$ and $B$ are defined as
\begin{equation}
A_1=\frac{1}{2g_0(3-\beta)M},\,\,\,\;\;\mbox{and}\,\,\,\,\;\,B=(3-\beta).
\end{equation}
Also here we have considered the positive sign for the solution given by Eq.(\ref{New6}). 

On the other hand, we get that  the number of $e$-folds between two different values of the scalar field $\phi_*$ and $\phi_f$ yields 
\begin{equation}
N=\int^{\phi_{*}}_{\phi_{f}}{\frac{1}{\sqrt{2\epsilon}}}d\phi=\frac{1}{4g_{0}^2M^2(3-\beta)^2(3+\beta)}\ln\Bigg[\frac{\big(-2\beta\sinh^2\Big(\frac{g_{0}(3-\beta)}{\sqrt{2}}M\phi\Big)+\beta+3\big)^{3(\beta+1)}}{\sinh^{2(\beta-3)}\Big(\frac{g_{0}(3-\beta)}{\sqrt{2}}M\phi\Big)}\Bigg]^{\phi_{*}}_{\phi_{f}},
\label{k32a}
\end{equation}
here we have considered  the first slow-roll parameter  given by Eq.(\ref{k29}). We emphasize  that the subscription $*$ is used to indicate the epoch in which the cosmological scale exits the horizon. In order to obtain analytical expressions for the cosmological perturbations,  we can assume for simplicity that  the value of the scalar field at the end of inflation $\phi_f$ satisfies the condition 
$\phi_f\gg\frac{\sqrt{2}}{g_{0}(3-\beta)M}$. Thus, from Eq.(\ref{k30}) we get that the value at the end of inflation of the scalar field is  reduced  to
\begin{equation}
\phi_f\simeq \frac{1}{\sqrt{2}g_0M(3-\beta)}\ln\left[\frac{3(\beta+1)}{\beta}\left(\frac{1-2g_0M(3-\beta)}{1-g_0M(3-\beta)}\right)\right].
\label{k312}
\end{equation}
By assuming the condition in which the scalar field $\phi\gg\frac{\sqrt{2}}{g_{0}(3-\beta)M}$, we find that the number of $e$-folds given by Eq.(\ref{k32a}) is simplified  to 
\begin{equation}
N\simeq\frac{1}{\sqrt{2}g_{0}M(3-\beta)}\Big(\phi_{*}-\phi_{f}\Big).
\label{NNNm}
\end{equation}

On the other hand, we can express
 the curvature spectrum $\mathcal{P}_{\mathcal{S}}$, the spectral index $n_{s}$ and the tensor to scalar ratio $r$ in terms of $N$ given that, this observational quantity should be evaluated when the cosmological scale exits the horizon \textit{i.e.}, 
 at $\phi=\phi_*$. Thus the power spectrum  from Eq.(\ref{17}) and considering Eqs.(\ref{k14}) and (\ref{k28}) we get 
 %}
\begin{equation}
\mathcal{P}_{\mathcal{S}}(\phi=\phi_*)=\frac{V_0[-2+B_1-2\cosh(2a\phi_*)]^3[\sinh(2a\phi_*)]^{-2}(2+3\sqrt{2}g_0H_0[\cosh(a\phi_*)]^{-2})^{5/2}}{192\sqrt{2}\,a^2\pi^2[1-B_1+\cosh(2a\phi_*)]^2[1+2\sqrt{2}\,g_0H_0(\cosh[a\phi_*])^{-2}]^{3/2}},
\end{equation}
or in terms of the number of $e-$folds $N$ we have
\begin{equation}
\mathcal{P}_{\mathcal{S}}(N)=\frac{V_0[B_1-4-4\mathcal{Q}^2]^3[4\mathcal{Q}^2(1+\mathcal{Q}^2)]^{-1}\left(2+\frac{3\sqrt{2}g_0H_0}{(1+\mathcal{Q}^2)}\right)^{5/2}}{192\sqrt{2}\,a^2\pi^2[2-B_1+2\mathcal{Q}^2]^2\left[1+\frac{2\sqrt{2}\,g_0H_0}{(1+\mathcal{Q}^2)}\right]^{3/2}},\label{k34}
\end{equation}
here the quantities $a$, $B_1$ and $\mathcal{Q}$ are
defined as
\begin{equation}
a=\frac{g_0(3-\beta)M}{\sqrt{2}},\,\,\,\,\,\,\,
B_1=\frac{B}{\beta}=\frac{3(1+\beta)}{\beta},
\label{z1}
\end{equation}
and 
\begin{equation}
\mathcal{Q}=\sinh\Big(\frac{g_{0}(3-\beta)}{\sqrt{2}}M\phi_{*}\Big)\simeq \frac{e^{g_0(3-\beta)M\phi_*/\sqrt{2}}}{2}=\frac{1}{2}\sqrt{\frac{3(1+\beta)(1+2g_0(\beta-3)M)}{\beta[1+ g_0(\beta-3)M]}}\,e^{[g_0(\beta-3)M]^2\,N}.
%,\hspace{1cm}\text{with}\hspace{1cm}\phi_{*}=\phi_%{f}+\sqrt{2}g_{0}M(3-\beta)N.
\label{k37}
\end{equation}
Here we have used that the value of the scalar field $\phi_*$ when the scale exits the horizon  is given by Eq.(\ref{NNNm}) and the value of $\phi$ at the end  of inflation i.e.,  $\phi_f$ is defined  by Eq.(\ref{k312}). Also in Eq.(\ref{k37}) we have considered  the approximation  $\phi\gg\frac{\sqrt{2}}{g_{0}(3-\beta)M}$.
%}

Also we can write the slow roll parameters $\epsilon$ and $\eta$ together with the function $\mathcal{A}$ as a function of the number of $e-$ folds $N$, in order to determine the scalar spectral index $n_s$ from Eq.(\ref{19}). Thus we have 
\begin{equation}
\epsilon=\frac{8a^2[B_1-2(1+\mathcal{Q}^2)]^2}{[B_1-4(1+\mathcal{Q}^2)]^2}\,\frac{\mathcal{Q}^2}{(1+\mathcal{Q}^2)},\,\,\,\,\,
\eta=4a^2\frac{\left(2+4B_1-4\mathcal{Q}^2-\frac{5B_1}{(1+\mathcal{Q}^2)}\right)}{[B_1-4(1+\mathcal{Q}^2)]},\,\,\,\,\mbox{and}\,\,\,\,
\mathcal{A}=\frac{3g_0H_0}{\sqrt{2}(1+\mathcal{Q}^2)}.
\label{z2}
\end{equation}

In this form, we find that the scalar spectral index in terms of the number of $e-$folds becomes
$$
n_s\simeq1-\frac{48a^2[B_1-2(1+\mathcal{Q}^2)]^2}{[B_1-4(1+\mathcal{Q}^2)]^2}\,\frac{\sqrt{2}\mathcal{Q}^2}{[\sqrt{2}(1+\mathcal{Q}^2)+3g_0H_0]}+8a^2\frac{\left(2+4B_1-4\mathcal{Q}^2-\frac{5B_1}{(1+\mathcal{Q}^2)}\right)}{[B_1-4(1+\mathcal{Q}^2)]}\frac{\sqrt{2}(1+\mathcal{Q}^2)}{[\sqrt{2}(1+\mathcal{Q}^2)+3g_0H_0]}\,
$$
\begin{equation}
-\,\frac{12\sqrt{2}aMg_0Q^2}{(1+2\sqrt{2}g_0H_0+Q^2)}\,\left(\frac{1+4\sqrt{2}g_0H_0+Q^2}{2+3\sqrt{2}g_0H_0+2Q^2}\right).\label{k35}
\end{equation}

Besides, we obtain that the tensor to scalar ratio $r=r(N)$   from Eq.(\ref{19b}) yields 
\begin{equation}
r=128\sqrt{2}a^2\,\mathcal{Q}^2\left(\frac{[2(1+\mathcal{Q}^2)-B]^2\,[(1+\mathcal{Q}^2)+2\sqrt{2}g_0H_0]^{3/2}}{(1+\mathcal{Q}^2)\,[B-4(1+\mathcal{Q}^2)]^2\,[2(1+\mathcal{Q}^2)+3\sqrt{2}g_0H_0]^{5/2}}\right).\label{k36}
\end{equation}

Now assuming the observational quantity 
$\mathcal{P}_{\mathcal{S}}$ and the scalar spectral index $n_s$   at the number of $e-$ folds $N=60$ \textit{i.e.} $\mathcal{P}_{\mathcal{S}}\simeq 10^{-9}$ and $n_s\simeq0.97$, and from Eqs.(\ref{k34}) and (\ref{k35}), we can obtain the parameters $M$ and $g_0$ using the upper and lower limits of parameter $3>\beta>0$. For the lower bound of $\beta$ and by fixing the value of $\beta=10^{-5}$, the parameters $M$ and $g_0$ are obtained $M\simeq1.1\times 10^{-6}$ and $g_0\simeq 1.3\times10^{5}$, respectively. Also from from Eq.(\ref{k36}), we find the tensor to scalar ratio $r$ as $r\sim 5\times 10^{-14}$. Now for the upper bound of $\beta$ \textit{i.e.} $\beta=2.99$, we find the parameters $M$ and $g_0$ as $M\simeq 1\times 10^{-5}$ and $g_0\simeq 4.4\times10^{6}$. 

From these values, we obtain the tensor to scalar ratio $r\sim 3\times 10^{-10}$. Thus, we find the observational constraints on $M$ and $g_0$ from the observational parameters considering $3>\beta>0$ are: $10^{-5}>M>10^{-6}$ and $4\times10^{6}>g_0>10^{5}$, respectively. Also, we note that for these values of the parameter-space of $\beta$, $M$ and $g_0$ the tensor to scalar ratio $r$ is very small ($r\sim 0$) and it 
 suggests a suppression of the gravitational wave for our model with  the coupling  $G(\varphi,\chi)$ given by Eq.(\ref{k1}). 

\subsubsection{Solution 2}

By assuming $C_{1}<0$ and $\frac{(6+\alpha)}{(3+\alpha)}>0$, the Hubble parameter (\ref{k8}) takes the following form
\begin{equation}
H(\varphi)=\frac{2\sqrt{2}g_{0}(6+\alpha)|C_{1}|}{(3+\alpha)}\frac{1}{\cos^{2}\Big(\frac{g_{0}(6+\alpha)(\varphi+C_{2})\sqrt{|C_{1}|}}{\sqrt{2}}\Big)}
\label{k18}    
\end{equation}
and then the potential (\ref{k10}) is obtained as
\begin{eqnarray}
V(\varphi)=\frac{96|C_{1}|^2g_{0}^2(6+\alpha)}{(3+\alpha)\cos^{4}\Big(\frac{g_{0}(6+\alpha)(\varphi+C_{2})\sqrt{|C_{1}|}}{\sqrt{2}}\Big)}\Bigg\{-\frac{3(2+\alpha)}{4(3+\alpha)}+\Big(1-\frac{(3+\alpha)}{12g_{0}^2|C_{1}|(6+\alpha)}\Big)\times\hspace{3cm}\nonumber\\
\times\cos^{2}\Big(\frac{g_{0}(6+\alpha)(\varphi+C_{2})\sqrt{|C_{1}|}}{\sqrt{2}}\Big)+\frac{(3+\alpha)}{12g_{0}^2|C_{1}|(6+\alpha)}\cos^{4}\Big(\frac{g_{0}(6+\alpha)(\varphi+C_{2})\sqrt{|C_{1}|}}{\sqrt{2}}\Big)\Bigg\}.\hspace{0.5cm}
\label{k19}    
\end{eqnarray}
From Eq.(\ref{k2}), we find the expression for $\dot{\varphi}$ as
\begin{equation}
\dot{\varphi}=-4\sqrt{|C_{1}|}\tan\Big(\frac{g_{0}(6+\alpha)(\varphi+C_{2})\sqrt{|C_{1}|}}{\sqrt{2}}\Big)
\label{k20}
\end{equation}
In the following, we work with a new scalar field $\phi=\varphi+C_{2}$ and $C_{1}=\pm M^2$. Then, the Hubble parameter (\ref{k18}) is driven as
\begin{equation}
H(\phi)=\frac{H_0}{\cos^2\Big(\frac{g_{0}(6+\alpha)}{\sqrt{2}}M\phi\Big)},\hspace{0.2cm}\text{where}\hspace{0.5cm}H_{0}=\frac{2\sqrt{2}g_{0}(6+\alpha)M^2}{(3+\alpha)}
\label{k21}    
\end{equation}
and the potential (\ref{k19}) can be found as
\begin{eqnarray}
V(\phi)=\frac{V_{0}}{\cos^{4}\Big(\frac{g_{0}(6+\alpha)}{\sqrt{2}}M\phi\Big)}\Bigg\{-\frac{3(2+\alpha)}{4(3+\alpha)}+\Big(1-\frac{(3+\alpha)}{12g_{0}^2M^2(6+\alpha)}\Big)\times\hspace{3cm}\nonumber\\
\times\cos^{2}\Big(\frac{g_{0}(6+\alpha)}{\sqrt{2}}M\phi\Big)+\frac{(3+\alpha)}{12g_{0}^2M^2(6+\alpha)}\cos^{4}\Big(\frac{g_{0}(6+\alpha)}{\sqrt{2}}M\phi\Big)\Bigg\}.\hspace{0.2cm}
\label{k22}    
\end{eqnarray}
where $V_{0}=\frac{96M^4g_{0}^2(6+\alpha)}{(3+\alpha)}$. Also, from eq.(\ref{k20}), evolution of the new scalar field $\phi$ versus cosmic time $t$ is found as
\begin{equation}
\phi(t)=\phi_0\sin^{-1}\Big(\exp\Big[-\frac{4g_{0}(6+\alpha)}{\sqrt{2}}M^{2}t\Big]\Big),\text{where}\hspace{0.5cm}\phi_{0}=\frac{\sqrt{2}}{g_{0}(6+\alpha)M}
\label{k23}    
\end{equation}
and then, combining the above relation with eq.(\ref{k21}), the Hubble parameter $H$ versus cosmic time can be obtained as
\begin{equation}
H(t)=\frac{H_0}{1-\exp\Big[-\frac{8g_{0}(6+\alpha)}{\sqrt{2}}M^{2}t\Big]}.
%\hspace{1cm}\mbox{and}\,\,\,\,\,\,a(t)\propto\Big(1-%e^{2(3+\alpha)H_0t}\Big)^{\frac{1}{2(3+\alpha)}}
\label{k24}    
\end{equation}
Here depending of the value of $\alpha$ assuming $g_0>0$, we can obtain two solutions for the scale factor $a(t)$ from Eq.(\ref{k24}). For the case in which the parameter $\alpha>-3$, we find that the scale factor as a function of the time yields
\begin{equation}
a(t)\propto\Big(1-e^{2(3+\alpha)H_0t}\Big)^{\frac{1}{2(3+\alpha)}},
\label{k25}
\end{equation}
and it corresponds to a negative scale factor for all time unless the power $(2(3+\alpha))^{-1}$ is even. However this situation can not occur and then it corresponds to a non physical solution.
In the case in which the parameter $\alpha<-6$, then the scale factor can be written as 
\begin{equation}
a(t)\propto\frac{e^{H_0t}}{\Big(1-e^{2|3+\alpha|H_0t}\Big)^{\frac{1}{2|3+\alpha|}}}
\label{k26}
\end{equation}
which is not again acceptable physically since it shows a negative value. 

\subsection{CASE III}

As one third case,  we consider the function $G(\varphi,\chi)$ depending to the both variables the scalar field $\varphi$ and its kinetic term $\chi$ as follows
\begin{equation}
G(\varphi,\chi)=\gamma_1+\gamma_2\varphi+\gamma_3\chi^{1/2},
\label{u1}
\end{equation}
where the quantities $\gamma_{1}$, $\gamma_{2}$ and $\gamma_{3}$ belong to different constants. For simplicity, we will consider positive quantities. From Eq.(\ref{n2}) and by setting $\gamma_2=1/2$ to obtain analytical solutions, then we get that the speed associated to scalar field becomes
\begin{equation}
\dot{\varphi}=-2\frac{H_\varphi}{AH},\hspace{0.5cm}\mbox{where}\hspace{0.5cm}A=\gamma_3(6+\alpha)/\sqrt{2}.
\label{u2}
\end{equation}
By derivation of the above expression with respect to the cosmic time $t$ and then by using the constant-roll condition (\ref{1}), we find the differential equation as follows
\begin{equation}
-(3+\alpha)H=-\frac{2}{A}\Big(\frac{H_{\varphi\varphi}H-H_{\varphi}^{2}}{H^{2}}\Big) .   
\label{u3}
\end{equation}
In order to solve this equation, we define a new variable $y=(2/A)\ln[AH]$, then Eq.(\ref{u3}) can be written as
\begin{equation}
y_{\varphi\varphi}=\frac{(3+\alpha)}{A}e^{Ay/2},  
\label{u4}
\end{equation}
which could be solved by
\begin{equation}
y(\varphi)=\frac{2}{A}\ln\Big[\frac{A^{2}C_{1}}{4(3+\alpha)}\frac{1}{\cosh^{2}\Big(\frac{A(\varphi+C_{2})\sqrt{C_{1}}}{4}\Big)}\Big],
\label{u5}
\end{equation}
where $C_{1}$ and $C_{2}$ are two integration constants. Thus, the Hubble parameter versus the scalar field
$\varphi$ yields
\begin{equation}
H(\varphi)=\frac{\gamma_3\,C_{1}(6+\alpha)}{4\sqrt{2}(3+\alpha)}\frac{1}{\cosh^{2}\Big(\frac{A(\varphi+C_{2})\sqrt{C_{1}}}{4}\Big)}.
\end{equation}
Here we note that the above  solution for Hubble parameter $H(\varphi)$ is similar to obtained in the case II in which the coupling function $G(\varphi,\chi)\propto\sqrt{\chi}$ (see Eq.(\ref{k1})), however the conditions for the parameter $\alpha$ and the integration constant $C_1$ are different.  Thus, we can have two solutions for the Hubble parameter $H(\varphi)$ depending to the sign of the integration constant $C_1$ and the ratio $(6+\alpha)/(3+\alpha)$. Unlike in the case II, two solutions of $H(\varphi)$ occur when: i) $C_1>0$ and $(6+\alpha)/(3+\alpha)>0$ or ii) $C_1<0$ and $(6+\alpha)/(3+\alpha)<0$.
%for $\gamma_{3}>0$ is given by
%\begin{equation}
%H(\varphi)=\frac{\gamma_{3}C_{1}(6+\alpha)}{4\sqrt{2}(3+\alpha)}\frac{1}{\cosh^{2}\Big(\frac{\gamma_{3}(6+\alpha)(\varphi+C_{2})\sqrt{C_{1}}}{4\sqrt{2}}\Big)}.
%\nonumber
%\end{equation}
In this context, we can write the two solutions for the Hubble parameter as a function of the scalar field as: 

\begin{equation}
  H(\varphi)=\begin{cases}
    \frac{\gamma_{3}C_{1}(6+\alpha)}{4\sqrt{2}(3+\alpha)}\frac{1}{\cosh^{2}\Big(\frac{\gamma_{3}(6+\alpha)(\varphi+C_{2})\sqrt{C_{1}}}{4\sqrt{2}}\Big)}, & \text{if $C_{1}>0$ and $\frac{(6+\alpha)}{(3+\alpha)}>0$ (with which $\alpha<-6$ or $\alpha>-3$)}.\\
    \frac{\gamma_{3}|C_{1}|(6+\alpha)}{4\sqrt{2}|3+\alpha|}\frac{1}{\cos^{2}\Big(\frac{\gamma_{3}(6+\alpha)(\varphi+C_{2})\sqrt{|C_{1}|}}{4\sqrt{2}}\Big)}, & \text{if $C_{1}<0$ and $\frac{(6+\alpha)}{(3+\alpha)}<0$\, (or $-6<\alpha<-3$)}.
  \end{cases}
  \label{u6}
  \end{equation}
  %\textcolor{blue}{
On the other hand, from the Friedmann equation (\ref{3}) and using Eq.(\ref{u2}), we find that the  effective potential can be written in terms of the Hubble parameters as
\begin{equation}
V(\varphi)=3H^{2}-\frac{12\sqrt{2}}{(6+\alpha)^{2}\gamma_{3}}\frac{H_{\varphi}^{2}}{H}.
\label{u7}    
\end{equation}
As before, in the following, we shall study two different forms of the Hubble parameters (\ref{u6}) with respect to the sign of the two parameters $C_{1}$ and $\alpha$. 

%through the condition $H>0$.

\subsubsection{Solution 1}

By considering $C_{1}>0$ and $\frac{(6+\alpha)}{(3+\alpha)}>0$, we can work with the first solution of the Hubble parameter (\ref{u6}) 
\begin{equation}
H(\varphi)= \frac{\gamma_{3}C_{1}(6+\alpha)}{4\sqrt{2}(3+\alpha)}\frac{1}{\cosh^{2}\Big(\frac{\gamma_{3}(6+\alpha)(\varphi+C_{2})\sqrt{C_{1}}}{4\sqrt{2}}\Big)}.
\label{u8}    
\end{equation}
Then, from Eq.(\ref{u7}), we find the reconstructed potential as a function of the scalar field by
\begin{equation}
V(\varphi)=\frac{3(6+\alpha)\gamma_{3}^{2}C_{1}^{2}}{8(3+
\alpha)\cosh^{4}\Big(\frac{\gamma_{3}(6+\alpha)(\varphi+C_{2})\sqrt{C_{1}}}{4\sqrt{2}}\Big)}\Bigg(\frac{18+5\alpha}{4(3+\alpha)}-\cosh^{2}\Big(\frac{\gamma_{3}(6+\alpha)(\varphi+C_{2})\sqrt{C_{1}}}{4\sqrt{2}}\Big)\Bigg).
\label{u9}    
\end{equation}
Here we note that the effective potential (\ref{u9}) is different with the potential obtained from the case II while both present a similar $H(\varphi)$.

Now, by plugging Eq.(\ref{u8}) into Eq.(\ref{u2}), we obtain that the speed of the scalar field by
\begin{equation}
\dot{\varphi}=\sqrt{C_{1}}\tanh\Big(\frac{\gamma_{3}(6+\alpha)(\varphi+C_{2})\sqrt{C_{1}}}{4\sqrt{2}}\Big),
\label{u10}    
\end{equation}
and this equation can be integrated to get $\varphi$ in terms of the cosmological time \textit{i.e.}, $\varphi(t)$.

%\textcolor{blue}{
In the following, we will assume that the integration constant $C_1$ is positive and equal to $C_{1}=M^{2}>0$ and introducing a new scalar field $\phi=\varphi+C_{2}$,  the Hubble parameter (\ref{u8}) versus the new scalar field $\phi$ takes the form
\begin{equation}
H(\phi)=\,\,\frac{H_0}{\cosh^{2}\Big(\frac{\gamma_{3}(6+\alpha)}{4\sqrt{2}}M\phi\Big)},\,\,\,\mbox{where}\,\,\,\,\,\,\,H_0=\frac{\gamma_{3}M^{2}(6+\alpha)}{4\sqrt{2}(3+\alpha)},
\label{u11}
\end{equation}
%Then, from eqs.(\ref{ne8}) and (\ref{ne9}), the %$\dot{\phi}$ and the potential are calculated as
%\begin{equation}
%\dot{\phi}=M\tanh\big[\frac{\gamma_{3}(6+\alpha)}{4\sqrt{2%}}M\phi\big],
%\label{ne11}    
%\end{equation}
and for the rebuilt effective potential (\ref{u9}), we have
\begin{equation}
V(\phi)=\frac{V_{0}}{\cosh^{4}\Big(\frac{\gamma_{3}(6+\alpha)}{4\sqrt{2}}M\phi\Big)}\Bigg(\frac{18+5\alpha}{4(3+\alpha)}-\cosh^{2}\Big(\frac{\gamma_{3}(6+\alpha)}{4\sqrt{2}}M\phi\Big)\Bigg),\hspace{0.5cm}\text{where}\hspace{0.5cm}V_{0}=\frac{3\gamma_{3}^{2}M^{4}(6+\alpha)}{8(3+
\alpha)}.
\label{u12}    
\end{equation}
In order to have an adequate reconstructed  effective potential \textit{i.e.} $V(\phi)>0$  %for the specific case in which  $(3+\alpha)>0$
, we need to consider the upper bound for the scalar field
\begin{equation}
\left[\frac{4\sqrt{2}}{\gamma_3(6+\alpha)M}\right]\cosh^{-1}\left[\sqrt{\frac{18+5\alpha}{4(3+\alpha)}}\right]>\phi.
\label{u13}
\end{equation}
 Here for simplicity, we have considered the positive sign of the square root. 
 %Also, from the above condition, one can %find that the only case $3+\alpha>0$ is %acceptable in order to escape from the %negative value of the scalar field. This %is compatible wit the condition used in %the Hubble parameter (\ref{u8}).} \\
%Hence, the potential (\ref{ne12}) can be rewritten as
%\begin{equation}
%V(\phi)=\frac{V_{0}}{\cosh^{4}\big[\frac{\gamma_{3}(3+|3+\alpha|)}{4\sqrt{2}}M\phi\big]}\Bigg(\frac{3+|3+\alpha|}{|3+\alpha|}-4\sinh^{2}\big[\frac{\gamma_{3}(3+|3+\alpha|)}{4\sqrt{2}}M\phi\big]\Bigg),\,\,\,\mbox{where}\,\,\,\,\,\,\,V_0=\frac{3(3+|3+\alpha|)\gamma_{3}^{2}M^{4}}{32|3+
%\alpha|}.
%\nonumber   
%\end{equation}

By solving Eq.(\ref{u10}) and then using Eq.(\ref{u11}), we find that the evolution of scalar field  in terms of the cosmological time results
\begin{equation}
\phi(t)=\phi_0\sinh^{-1}\Big(\exp\Big[\frac{\gamma_{3}(6+\alpha)}{4\sqrt{2}}M^{2}t\Big]\Big)+C_0,\,\,\,\,\,\,
%\mbox{and}\hspace{1cm}H(t)=\frac{H_0}
%%{1+\exp\Big[\frac{\gamma_{3}(6+\alpha)}
%%{2\sqrt{2}}M^{2}t\Big]},
\label{u14}    
\end{equation}
 where we have defined 
 $\phi_0=\frac{4\sqrt{2}}{M(6+\alpha)\gamma_{3}}$ and $C_0$ corresponds to the integration constant. Using the definition $H\equiv\frac{\dot{a}}{a}$, we find that the evolution of the scale factor versus cosmic time for $\alpha>-3$ 
 becomes similar to Eq.(\ref{k17}). Then, we have an inflationary universe since $\ddot{a}>0$. Notice that the solution of the model for the equivalent case $\alpha<-6$ (see eq.(\ref{u6})) also shows an inflationary behaviour ($\ddot{a}>0$). 
 %takes the following from
%\begin{equation}
%a(t)\propto \frac{e^{H_0t}}%{\Big(1+e^{2(3+\alpha)H_0t}\Big)^{\frac{1}
%{2(3+\alpha)}}}.
%\label{u15}    
%\end{equation}
%This scale factor shows that the solution 
%(\ref{u8}) is an inflationary solution %because of  $\ddot{a}>0$.
%\begin{figure*}[!hbtp]
	%\centering
    %\includegraphics[width=.40\textwidth,keepaspectratio]{u1.pdf}
	%\hspace{0.5cm}
	%\includegraphics[width=.32
	%\textwidth,keepaspectratio]{new2.pdf}
	%\includegraphics[width=.32\textwidth,keepaspectratio]{new3.pdf}
	%\hspace{0.5cm}
	%\includegraphics[width=.32
	%\textwidth,keepaspectratio]{new4.pdf}
	%\caption{The behaviour of potential (\ref{u12}) versus the new scalar field $\phi$ for $\alpha=-0.5$.}
	%\label{fignew1}
%\end{figure*}\\
%\textcolor{blue}{

In the following, we perform the  analysis of the cosmological perturbations using the potential of the model for  rewritten in terms of $\beta=-(3+\alpha)$. Here if $\alpha<-6$ then $\beta>3$ and if $\alpha>-3$ then the parameter $\beta$ is negative ($\beta<0$). Thus introducing the parameter $\beta$ the reconstruct effective potential becomes
\begin{equation}
V(\phi)=\frac{V_{0}}{\cosh^{4}\Big(\frac{\gamma_{3}(3-\beta)}{4\sqrt{2}}M\phi\Big)}\Bigg(\frac{5\beta-3}{4\beta}-\cosh^{2}\Big(\frac{\gamma_{3}(3-\beta)}{4\sqrt{2}}M\phi\Big)\Bigg),\hspace{0.5cm}\text{where}\hspace{0.5cm}V_{0}=\frac{3\gamma_{3}^{2}M^{4}(\beta-3)}{8\beta}.
\label{u18}    
\end{equation}

 Further the slow roll parameter associated to the Hubble parameter $\epsilon_1$ defined in Eq.(\ref{7}) becomes 
%\textcolor{blue}{
\begin{equation}
\epsilon_{1}=-\frac{\dot{H}}{H^2}=\frac{2mM}{H_{0}}\sinh^{2}[m\phi],
%\hspace{0.5cm}\eta_{H}=-%\frac{\ddot{H}}%%%%
%
%{2H\dot{H}}=\frac{mM}%{H_{0}}\big(\cosh^2[m\phi]-2\big)
\end{equation}
where the quantity $m$ is defined as $m=\frac{\gamma_3(3-\beta)M}{4\sqrt{2}}$.
%and $b_1=\frac{5\beta-3}%
%
%{4\beta}$.

Thus, by setting $\epsilon_{1}=1$ or equivalently $\ddot{a}=0$,  we obtain that the value of inflaton field at the end of inflation is
\begin{equation}
\phi_{f}=\frac{1}{m}\arcsinh\Big(\sqrt{\frac{H_{0}}{2mM}}\Big).
\label{N1}
\end{equation}
%}

For the number of $e-$ folds $N$, using the reconstructed Hubble parameter (\ref{u11}) and the expression of $\dot{\phi}$ given by Eq.(\ref{u10}) we have
\begin{equation}
N=\int^{\phi_{f}}_{\phi_{*}}{Hdt}=\frac{H_{0}}{mM}\ln\Big(\tanh[m\phi]\Big)\bigg|^{\phi_{f}}_{\phi_{*}}.
\end{equation}
Thus, we find that the value of the scalar field when the scale exits the horizon can be written as

\begin{equation}
\phi_{*}=\frac{1}{m}\arctanh\Big[\frac{q}{\sqrt{q^2+1}}e^{-\frac{mMN}{H_{0}}}\Big],\hspace{0.5cm}\text{with}\hspace{0.5cm}q\equiv\sqrt{\frac{H_{0}}{2mM}}.
\end{equation}

On the other hand, in order to analyze the cosmological perturbation,
 we determine the parameter $\varepsilon_s$ and $c_s$ from Eq.(\ref{14}). Thus we find that   these parameters become
\begin{equation}
\varepsilon_s=\frac{\sqrt{2}\gamma_3M^2\sinh^2[m\phi]}{H_0}\,,
\end{equation}
and 
\begin{equation}
c_s^2=\gamma_3\left[\frac{(4H_0 \sech^2[m\phi]-\sqrt{2}\gamma_3M^2\tanh^{2}[m\phi])}{4\sqrt{2}+3\sqrt{2}\gamma_3^2M^2\tanh^2[m\phi]+24\gamma_3H_0\sech^2[m\phi]-6H_0M^2\sech^2[m\phi]\tanh^2[m\phi]}\right].
\label{z4}
\end{equation}
From Eq.(\ref{13}) we obtain that the power spectrum becomes
$$
\mathcal{P}_S(\phi=\phi_*)=\left(\frac{H_0^3}{8\sqrt{2}\pi^2\gamma_3^{3/2}M^2\cosh^2[m\phi_*]\sinh^2[m\phi_*]}\right)\,\times
$$
\begin{equation}
\times\left[\frac{4\sqrt{2}+3\sqrt{2}\gamma_3^2M^2\tanh^2[m\phi_*]+24\gamma_3H_0\sech^2[m\phi_*]-6H_0M^2\sech^2[m\phi_*]\tanh^2[m\phi_*]}{(4H_0 \sech^2[m\phi_*]-\sqrt{2}\gamma_3M^2\tanh^{2}[m\phi_*])}\right]^{1/2}.\label{P3}
\end{equation}
For the spectral scalar index $n_s$ from Eq.(\ref{ns2}) yields
$$
n_s=1-2\frac{\sqrt{2}\gamma_3M^2\sinh^2[m\phi_*]}{H_0}-\frac{2mM\cosh^2[m\phi_*]}{H_0}\,+
$$
\begin{equation}
+\frac{mM\,\tanh^2[m\phi_*]}{H_0}\,\left(\frac{(b(d_1+d_2)+c(d_1+d_3-d_4)+d_4(c\,\cosh[2m\phi_*]-b)\sech^4[m\phi_*])}{[(b+c)\sech^2[m\phi_*]-c][d_1+d_2-(d_2-d_3+d_4)\sech^2[m\phi_*]+d_4\sech^4[m\phi_*]}\right),\label{ns3}
\end{equation}
where
\begin{equation}
b=4H_0,\,\,\,\,\,\,c=\sqrt{2}\gamma_3M^2,\,\,\,d_1=4\sqrt{2},\,\,\,\,\,\,d_2=3\sqrt{2}\gamma_3^2M^2,\,\,\,\,\,\,d_3=24\gamma_3H_0,\,\,\mbox{and}\,\,\,\,\,\,d_4=6H_0M^2.
\label{z5}
\end{equation}
Also, we find that the tensor to scalar ratio $r$ in terms of the scalar field when the scale exists the horizon results
$$
r=16\varepsilon_c\,c_s=16\frac{\sqrt{2}\gamma_3^{3/2}M^2\sinh^2[m\phi_*]}{H_0}\,\times
$$
\begin{equation}
\times\left[\frac{(4H_0 \sech^2[m\phi_*]-\sqrt{2}\gamma_3M^2\tanh^{2}[m\phi_*])}{4\sqrt{2}+3\sqrt{2}\gamma_3^2M^2\tanh^2[m\phi_*]+24\gamma_3H_0\sech^2[m\phi_*]-6H_0M^2\sech^2[m\phi_*]\tanh^2[m\phi_*]}\right]^{1/2}.
\end{equation}

From these results, we can obtain the constraints for the parameters $\gamma_3$ and $M$, when the number of \textit{e}-folds $N$, the parameter $\beta$, the scalar power spectrum and the scalar spectral index are given. From Eqs.(\ref{P3}) and (\ref{ns3}), for the negative values of $\beta$, $N=60$, the power spectral $\mathcal{P}_s\simeq10^{-9}$ and the scalar index $n_s\simeq 0.97$, we find the constraints on parameters $\gamma_3$ and $M$ as  $\gamma_3\simeq2\times10^{-4}$ and $M=5\times10^{-3}$ where we have fixed the small negative value $\beta=-0.01$. 

In the situation in which we have used  positive values for $\beta> 3$ or negative values such that  $\beta<-0.01$ and $-0.009<\beta<0$, we do not find solutions for $\gamma_3$ and $M$ from the observational data.  For the values obtained of the parameters  ($\beta,\gamma_3,M$), we find that tensor to scalar ratio $r$ is very small in which $r\sim10^{-6}$ and it suggests a suppression of the gravitational wave for the coupling function $G(\varphi,\chi)$ given by Eq.(\ref{u1}). For positive values of $\beta>3$, we do not find solutions for the parameter-space from the system  (\ref{P3}) and (\ref{ns3}), with which for $\beta>3$ the model does not work. 

\subsubsection{Solution 2}

In the case $C_{1}<0$ and $\frac{(6+\alpha)}{(3+\alpha)}<0$, we get the Hubble parameter (\ref{u6}) 
\begin{equation}
H(\varphi)=\frac{\gamma_{3}|C_{1}|(6+\alpha)}{4\sqrt{2}|3+\alpha|}\frac{1}{\cos^{2}\Big(\frac{\gamma_{3}(6+\alpha)(\varphi+C_{2})\sqrt{|C_{1}|}}{4\sqrt{2}}\Big)},
\label{u16}    
\end{equation}
and also from eq.(\ref{u7}), the effective potential takes the form
\begin{equation}
V(\varphi)=\frac{3(6+\alpha)\gamma_{3}^{2}|C_{1}|^{2}}{8|3+
\alpha|\cos^{4}\Big(\frac{\gamma_{3}(6+\alpha)(\varphi+C_{2})\sqrt{|C_{1}|}}{4\sqrt{2}}\Big)}\Bigg(\cos^{2}\Big(\frac{\gamma_{3}(6+\alpha)(\varphi+C_{2})\sqrt{|C_{1}|}}{4\sqrt{2}}\Big)-1+\frac{6+\alpha}{4|3+\alpha|}\Bigg).
\label{u17}    
\end{equation}
However as the Hubble parameter in terms of the scalar field is equivalent to the Hubble parameter analyzed in the case II, (see Eq.(\ref{k21})),   we conclude that this model does not work if we consider that the parameters $C_1$ and $(6+\alpha)/(3+\alpha)$ are negative.

\section{Conclusion}\label{Con}

In this paper, we have investigated different constant roll models, in the context of the Galilean inflation. In the constant roll consideration we have found a general relationship between the function  associated with the Galileon term $G(\varphi,\chi)$ and the  Hubble parameter and its derivatives  in terms  of  the scalar field. This led us to a general criterion for the reconstruction of the background variables, in particular, the effective potential and the Hubble parameter in terms of the scalar field.

In order to present an analytical discussion in the context of the constant-roll approach, we have considered that these models are described by different powers of the parameters $\nu$ and $n$ associated with the function  $G(\varphi,\chi)\propto\varphi^\nu\,\chi^n$. Also, we have studied the situation in which the coupling function $G(\varphi,\chi)$ depends of both variables; the scalar field $\varphi$ and its kinetic energy $\chi$, see Eq.(\ref{u1}).

For the specific case in which the coupling function $G(\varphi,\chi)\propto\varphi^\nu\,\chi^n$, as a
 first case, we took $\nu=1$ and $n=0$ wherewith the function $G(\varphi,\chi)=\gamma\varphi$, where $\gamma$ corresponds to a constant. Here, we have found the reconstruction for the Hubble parameter and the effective potential in terms of the scalar field together with the solution for the scalar field as a function of the time \textit{i.e.}, $\phi(t)$. Also, in the particular case in which the $\beta=-(3+\alpha)>0$ and the parameter $\gamma<1/2$, we have obtained the expressions for the curvature power spectrum, scalar spectral index and the tensor-to-scalar ratio in terms of the number of $e-$folds $N$. In this way, using the Planck data and for the specific case in which the number of the $e-$folds $N=60$, we have found the constraints on the parameters  $M$, $\alpha$ and $\gamma$. In this respect, we have obtained a  narrow range for the parameters $\gamma\sim 1/2$ and $\beta\sim{\mathcal{O}}(1)$, however for the energy scale $M$, we have found that in order to satisfy the constraints from Planck data the range for $M$ becomes  $10^{-5}>M>10^{-11}$.
In Fig.\ref{fig1AA} we show the contours curve related to the scalar spectral index as well to the tensor to scalar ratio for different values of the $\beta$ and $\gamma$ considering Eqs.(\ref{n15}) and (\ref{r1}), respectively. Here in both panels we have considered the number of $e$-folds $N=60$. From these plots, given a value of $n_s$ or $r$, one can constraint and visualize the parameter space ($\beta$,$\gamma$) from the observational Planck data.

In the second case in which the function $G(\varphi,\chi)\propto\varphi^\nu\,\chi^n$, we have considered the specific situation in which $\nu=0$ and $n=1/2$ and then the function $G(\varphi,\chi)$ is defined as $G(\varphi,\chi)=g_0\sqrt{\chi}$, which $g_0$ a new constant. In this case, we have analyzed two situations in which the integration constant is positive $C_1>$ and when  $C_1<0$, together with a specific range associated with the parameter $\alpha$ (or equivalently $\beta$).  In the situation in which the integration constant $C_1>0$ and the range for $-6<\alpha<-3$ (solution 1), we have obtained the reconstruction of  
the effective potential $V(\varphi)$ together with the Hubble parameter  and also we have shown that in this case the scale factor presents an accelerated expansion ($\ddot{a}>0$). In the context of the cosmological perturbations, we have found the  spectrum $\mathcal{P}_{\mathcal{S}}$, the index $n_s$ and the ratio $r$ as a function of the number $N$. As before, by considering the Planck data together with the number of $e-$folds $N=60$ and assuming the range for $\beta$ given by $3>\beta>0$, we have obtained the constraints on the parameter-space for our model  (solution 1). In this form,  we have found that the range for the energy scale $M$ as well for the parameter $g_0$ are; $10^{-5}>M>10^{-6}$ and $10^{6}>g_0>10^{5}$ for $3>\beta>0$. From these values, we have determined that the tensor to scalar ratio, is very small  and then the coupling $G(\varphi,\chi)\propto \sqrt{\chi}$ predicts a ratio $r\sim 0$. In the case in which the integration constant $C_1<0$ and $\alpha>-3$ or $\alpha<-6$ (solution 2), we have found that the solution for the scale factor shows a non inflationary behaviours ($\ddot{a}<0$) and then the model does not work to 
describe an inflationary epoch.

For the case III in which the coupling function is given by Eq.(\ref{u1}) and it depends of both variables, we have studied two cases in which the integration constant $C_1$ is positive and also when the  constant $C_1$ is negative
together with a special range of the parameter $\alpha$ (or $\beta$). For the case in which the integration constant $C_1>0$ and the parameter $\beta$ satisfies $\beta<0$ or $\beta>3$ (solution 1), we have obtained the reconstruction of the variables of background and a scenario inflationary. In relation to the cosmological perturbations, we have determined the power spectrum, scalar index and the tensor to scalar ratio under the slow roll approximation. Here only for a small negative value of $\beta$ closest to zero ($\beta=-0.01$) we have found appropriate values of $\gamma_3$ and the energy scale $M$ in order to satisfy the Planck data. From these constraints, we have obtained that the tensor to scalar ratio  $r\sim0$ and then the coupling $G(\varphi,\chi)$ of the case III gives account of  a suppression of the gravitational wave. 
In the case in which the parameter $\beta$ is positive ($\beta>3$) or negative values such that  $\beta<-0.01$ and $-0.009<\beta<0$, numerically we have found that the model does not work from observational data (Planck). For the solution 2 in which the integration constant $C_1$ is negative  and the range for the parameter $-6<\alpha<-3$, we have determined that the model does not work, since the solution for the scale factor is negative and then it corresponds to a non-physical solution. 

Finally, in this article we have not addressed another functions $G(\varphi,\chi)$ to study the model of constant roll G-inflation due to difficulty of obtaining analytical solutions to the background variables. Also, we have not addressed  our research on the incorporation of non-canonical K-inflation terms to establish its contribution to the Galileon term in our model. In relation to the cosmological perturbations,   we have not focused in this article to analyze the modification of  scalar power spectrum (super Hubble evolution) in the context of the Galilean constant-roll scenario, see     Ref.\cite{Motohashi9}.   
We hope to return to address these points in the framework of constant roll G-inflation in the near future.

\bibliographystyle{ieeetr}
\bibliography{biblo}

\end{document}